\documentclass[a4paper,11pt]{article}
\pdfoutput=1 % if your are submitting a pdflatex (i.e. if you have
\usepackage{jheppub} % for details on the use of the package, please
\usepackage[T1]{fontenc} % if needed

\usepackage{url}

\usepackage{tikz}
\usepackage{youngtab}
\usepackage{booktabs}
\usepackage{cellspace}
\usepackage{makecell}
\setcellgapes{5pt}
\usepackage{multirow}
\usepackage{tabularx}
\usepackage{nicefrac}
\usepackage{latexsym,amsmath,amsfonts,amssymb,mathrsfs,bm,empheq,bbm}

  {\left\lbrace\begin{array}{@{}l@{}}}%
  {\end{array}\right.}

\title{\textbf{Multicritical hypercubic models}}

\author[a]{R.\ Ben Al\`i Zinati,}%\note{Corresponding author.}}
\author[b]{A.\ Codello}
\author[c]{and O.\ Zanusso}

\affiliation[a]{Sorbonne Universit\'e \& CNRS, Laboratoire de Physique Th\'eorique de la Mati\`ere Condens\'ee, LPTMC, F-75005,  Paris, France}
\affiliation[b]{Instituto de F\'isica, Faculdad de Ingenier\'ia, Universidad de la Rep\'ublica,\\ 11000 Montevideo, Uruguay}
\affiliation[c]{Universit\`a di Pisa and INFN - Sezione di Pisa, Largo Bruno Pontecorvo 3, I-56127 Pisa, Italy}

\emailAdd{riccardo.ben\_ali\_zinati@sorbonne-universite.fr}
\emailAdd{acodello@fing.edu.uy}
\emailAdd{omar.zanusso@unipi.it}

\abstract{
We study renormalization group multicritical fixed points in the $\epsilon$-expansion of scalar field theories characterized by the symmetry of the (hyper)cubic point group $H_N$.
After reviewing the algebra of $H_N$-invariant polynomials and arguing that there can be an entire family of multicritical (hyper)cubic solutions
with $\phi^{2n}$ interactions in $d=\frac{2n}{n-1}-\epsilon$ dimensions,
we use the general multicomponent beta functionals formalism to study the special cases $d = 3-\epsilon$ and $d =\frac{8}{3}-\epsilon$,
deriving explicitly the beta functions describing the flow of three- and four-critical (hyper)cubic models.
We perform a study of their fixed points, critical exponents and quadratic deformations for various values of $N$, including
the limit $N=0$, that was reported in another paper in relation to the randomly diluted single-spin models,
and an analysis of the large $N$ limit, which turns out to be particularly interesting since it depends on the specific multicriticality.
We see that, in general, the continuation in $N$ of the random solutions is different from the continuation coming from large-$N$,
and only the latter interpolates with the physically interesting cases of low-$N$ such as $N=3$.
Finally, we also include an analysis of a theory with quintic interactions in $d =\frac{10}{3}-\epsilon$ and,
for completeness, the NNLO computations in $d=4-\epsilon$.}

\begin{document}
\maketitle

\section{Introduction}
\label{sec:intro}

The original motivations for considering the cubic point group and its generalization, the hypercubic point group $H_N$,
in statistical mechanics was to describe the critical properties of classical three-dimensional spin systems
in which the interactions are modulated by an underlying cubic lattice~\cite{Aharony:1973b, Wallace:1973}.
In some magnets, like for example Fe or Ni, ions are arranged in crystals with cubic symmetry:
non-rotationally invariant single-ion interactions such as $\sum_{i,\alpha} (\sigma_i^{\alpha})^4$
($\sigma_i^\alpha$ is the $i$-th component of the spin vector $\bm{\sigma}$ at the lattice point $\alpha$)
are allowed alongside the ``standard'' isotropic term $\sum_{i} (|\bm{\sigma}_i|^2)^2$.
The inclusion of anisotropic terms in the $N$-component isotropic Ginzburg-Landau (GL) Hamiltonian
changes the critical behavior from the $O(N)$-invariant isotropic one
to a distinctive hypercubic critical behavior provided that $N>N_c\simeq 2.9$ in $d=3$ dimensions.
Using the renormalization group (RG) approach to critical phenomena, this can be seen from the fact that the hypercubic critical point
is more infrared relevant than the isotropic one for values of $N$ above $N_c$, which therefore includes
the physically relevant case of $N=3$ \cite{Carmona:1999, Adzhemyan:2019gvv}.
Given that $N=3$ is just barely above $N_c$, the case of cubic anisotropy is of paradigmatic importance
in exemplifying how the symmetry of a system influences the nature of its phase transition.

Another important reason has to do with the inclusion of random disorder in traditional magnetic systems
with $\mathbb{Z}_2$ symmetry.
For sufficiently low concentrations of non-magnetic impurities, such as Zn atoms in a Fe based Ising-like lattice,
a ferromagnetic system still exhibits a second-order phase transition, but its critical properties are different from those of the pure system \cite{Harris_1974}.
To include the effect of disorder in the GL description one usually resorts to the replica approach,
in which $N$ identical copies of the system are introduced. Upon integration over the distribution of the disorder,
the system has an enhanced $H_N$ symmetry and it is possible to show that the critical exponent
of the quenched averages of correlators can be obtained from the limit $N\to 0$ \cite{cardy1996scaling}.
Using again the RG approach, it is possible to see that the replicated theory admits
both a decoupled FP, representing $N$ copies of the original pure system,
and a new random FP with true $H_N$ symmetry, representing the diluted system.
Similarly to the previous motivation, the random FP is more infrared relevant than the decoupled one \cite{Pelissetto:2000ek}.

The above discussions are implicitly based on a GL Hamiltonian with $\phi^4$-like interactions, but,
in principle, higher-order interaction terms of the form $\phi^{2n}$ can also be included in the GL description, both with isotropic and $H_N$ symmetry \cite{Zinati:2019gct,henriksson2020analytic}.
These are generally discarded, at least in qualitative approaches,
by following standard considerations based on the canonical dimension of the operators
that they would correspond to in the GL Hamiltonian for a three-dimensional system.
Higher-order operators, however, play an important role in lower dimensional systems,
because they give rise to multicritical behaviors,
given that they become increasingly RG relevant when lowering the dimensionality of the system.
Rather generally, by means of a standard Hubbard-Stratonovich transformation that introduces a scalar $N$-component field variable $\phi_i$,
microscopic terms like $\sum_{i,\alpha} (\sigma_i^{\alpha}\sigma_i^{\alpha})^{n}$, representing a higher-order isotropic interaction,
and $\sum_{i,\alpha} (\sigma_i^{\alpha})^{2n}$, representing a higher-order single-ion interaction, for $n\geq 3$
can be expressed as $\sum_{i}(\phi_i\phi_i)^n$ and $\sum_{i}\phi_i^{2n}$, respectively.
A GL Hamiltonian based on these $\phi^{2n}$-like interactions can be renormalized perturbatively in $d_c=\frac{2n}{n-1}$ dimensions,
and its renormalization gives the critical properties of the multicritical phase transitions in terms of an $\epsilon$-expansion
in $d=d_c-\epsilon$ dimensions \cite{itzykson2012quantum}. Notice also how the textbook example of $\phi^4$ is recovered for $n=2$.

Multicritical models are important for both theoretical and experimental reasons.
Two well-known examples of multicritical behaviors based on the isotropic model with $\phi^6$-interaction
are provided by the tricritical point of He$^3$-He$^4$ mixtures,
which is microscopically described by the so-called Blume-Capel model \cite{GriffithsBlumeCapel, BlumeEmeryGriffiths, RiedelWegner},
and the Flory $\theta$-point of very long polymer chains,
which is related to the limit of $N\to 0$ components \cite{DeGennes1972, DeGennes1975, Duplantier1982}.
On the more theoretical side, single component's $\phi^{2n}$ models with $\mathbb{Z}_2$ symmetry can be seen as multicritical generalizations
of the Ising universality class and are famously connected to the infinite set of unitary minimal models $\mathcal{M}_{n+1,n+2}$,
that are conformal field theories (CFTs)
emerging from the representations of the infinite-dimensional conformal group in $d=2$ \cite{zamolodchikov:1986}.
It is possible, though still unknown for the most part, that more complicated symmetry groups, such as the aforementioned $O(N)$ and $H_N$,
admit the same type of multicritical generalization of the Ising case and are also CFTs with a special meaning in $d=2$ \cite{Wiese2,Wiese1,henriksson2020analytic}.

In this paper, we move towards a better understanding of the last paragraph
by analyzing the renormalization group fixed points of multicritical scalar field theories obeying the symmetry of the hypercube.
The paper is structured as follows:
In Sect.~\ref{sec:hypercubicalgebra}, we outline the algebra of hypercubic invariant polynomials,
which allows us to determine the basis of invariant polynomials needed in the construction of the GL Hamiltonian.
In Sect.~\ref{sec:quadraticdeformations}, we discuss the representation theory of the hypercubic group in relation
to quadratic composite operators, because they have a pivotal role in discussing the symmetry breaking patterns
connecting various critical points.
In Sect.~\ref{sec:fprg}, we review the main points behind the perturbative RG method that we use to obtain the $\epsilon$-expansion.
Sections~\ref{sec:3-eps} and \ref{sec:8/3-eps} contain our results on the multicritical generalizations of the hypercubic model
with $\phi^6$- and $\phi^8$-like interactions, which are the main results of this paper.
Conclusion and perspectives are given in Section~\ref{sec:conclusions}.
We also include appendix~\ref{sec:4-eps}, which summarizes several results for the hypercubic theory in $d=4-\epsilon$ dimensions for completeness,
and appendix~\ref{sec:10/3-eps}, which presents a non-unitary generalization of the hypercubic model in $d=\frac{10}{3}-\epsilon$ dimensions.

\section{Algebra of hypercubic invariant polynomials}
\label{sec:hypercubicalgebra}

We begin by reviewing some fundamental facts on
the algebra of invariant polynomials of the hypercubic point group $H_N$,
which we need to construct a basis for the interaction potentials.
For this purpose, let $\mathcal{R}$ be the algebra of all polynomials $f$ in the variables $\{\phi_1, \dots, \phi_N\}$
with coefficients in $\mathbb{C}$.
The hypercubic group acts on $\{\phi_1, \dots, \phi_N\}$ as the subgroup of $O(N)$ that leaves a unit hypercube
centered at the origin of $\mathbb{R}^N$ invariant. The group can be generated from all permutations of the axes,
and a parity for each axis, resulting in the isomorphism $H_N\simeq(\mathbb{Z}_2)^N\rtimes S_N$, where $S_N$ is the permutation
group of $N$ objects.
We are interested in the subalgebra $\mathcal{R}^{H_N}$ of all the polynomials $f \in \mathcal{R}$ satisfying ${g} f = f$ for
each ${g} \in H_N$.

It is important to notice two things.
First, the algebra
 $\mathcal{R}^{H_N}$ admits the decomposition
$\mathcal{R}^{H_N} = \bigoplus_k \mathcal{R}^{H_N}_k$, where $\mathcal{R}^{H_N}_k$ consists of all the invariant homogeneous polynomials of degree $k$,
in the usual sense. Hence, in order to determine $\mathcal{R}^{H_N}$,
 it is sufficient to determine each $\mathcal{R}^{H_N}_k$. Second, we can introduce a generating function for the dimension of the homogeneous component $\mathcal{R}^{H_N}_k$. This is a
formal power series in a variable $t$, known as the Hilbert series (also known as
the Poincar\'e series, or the Molien series) of the ring $\mathcal{R}^{H_N}$
\begin{equation}
  M(t) \equiv \sum_{k \geq 0} (\dim \mathcal{R}^{H_N}_k)~t^k \,.
\end{equation}
The Hilbert series can be determined explicitly for the ring of any finite group through Molien's theorem. In the case of $H_N$, one finds
\begin{equation}
          M(t) = [(1-t^2)(1-t^4)(1-t^6)(1-t^8)\dots(1-t^{2N})]^{-1} \,.
\end{equation}
This result is very useful for us since it determines the size of the ring $\mathcal{R}^{H_N}$
and, therefore, the number of couplings that we need to parametrize the potential.
Once the series is expanded,
the coefficient of the monomial $t^{2n}$ gives the dimension of the corresponding subspace $\mathcal{R}^{H_N}_{2n}$.
We summarize the first few values of $\dim \mathcal{R}^{H_N}_k$ in a table.
{\makegapedcells
\begin{center}
\begin{tabular}{ |c|c|c|c|c|c|c|c|c|c| }
 \hline
 $\mathcal{R}^{H_N}_2$ & $\mathcal{R}^{H_N}_4$ & $\mathcal{R}^{H_N}_6$ & $\mathcal{R}^{H_N}_8$ &
 $\mathcal{R}^{H_N}_{10}$ & $\mathcal{R}^{H_N}_{12}$ & $\mathcal{R}^{H_N}_{14}$ & \dots & $\mathcal{R}^{H_N}_{2n}$ \\
 $1$ & $2$ & $3$ & $5$ & $7$ & $11$ & $15$ & \dots & $p(n)$ \\
 \hline
\end{tabular}
\end{center}
}

The counting rule above applies, to some extent, rather generally, so it can be used to continue
our results analytically in $N$, and eventually it is responsible for the determination of $N$-dependent renormalization group functions.
It is important to stress, however, that the counting rule of the number of operators applies only for $N$ sufficiently big
in comparison to the order of the interaction $2n$, that is, $N \geq n$.
If instead we concentrate on the natural values for which $N<n$, then the number of independent monomials (and therefore of associated coupling constants) is actually lower, because one or mode interactions of the general basis can be expressed
in terms of the others. In these cases, which require separate analyses given in the later parts of Sections \ref{sec:3-eps} and \ref{sec:8/3-eps},
the system of beta functions is reduced in a specific way that accounts for the degeneracy of linear combinations of the operators.

We find most convenient to work always with the general case, and eventually specialize the results later.
In this way, one can show that the number of invariant polynomials of degree $2n$ corresponds to the number $p(n)$ of partitions of the positive integer $n$.
To clarify this relation, we first define generalizations of the Kronecker delta,
which are symmetric tensors with $2n$ indices defined as
\begin{equation}\label{eq:deltatensors}
    \delta_{i_1 \dots i_{2n}} =
    \begin{cases}
    1\,,~~\text{when} ~ {i_1=i_2=\dots= i_{2n}}\\
    0\,,~~\text{otherwise}\,.
    \end{cases}
\end{equation}
Homogeneous invariant polynomials of degree $2n$ in the variables $\phi_{i}$
can be obtained by considering all possible ways to contract the product $\phi_{i_1} \dots \phi_{i_{2n}}$
with combinations of the basic tensors $\delta_{i_1 \dots i_{2k}}$ with $k\leq n$.
This can be done exactly in $p(n)$ ways, as expected from the Molien series.
A basis made with all possible contractions is manifestly invariant under $H_N$, in fact
the hypercubic group acts by reflecting and permuting the variables $\phi_i$,
namely through the transformations $\phi_i \to  -\phi_i$ and $\phi_i \,\to\, \phi_j$ for all $i$ and $j$,
thanks to the fact that $\delta_{i_1 \dots i_{2n}}$ is a symmetric tensor with an even number of indices.

\section{Quadratic deformations}\label{sec:quadraticdeformations}

In this section, we consider the study of non-singlet quadratic operators since they turn out to be particularly important in the determination of the so-called crossover exponents \cite{AHARONY1976}.
On general grounds, there are physical situations in which the critical behavior depends on the interplay of more than one order parameter, or, alternatively, on different components of a multicomponent order parameter.
In these cases, the lattice Hamiltonian typically presents a quadratic term, weighted by some parameter $g$, which reflects weak anisotropies between such competing order parameters.
By approaching the critical temperature $m_c(g)$, the isotropic transition crosses over to a transition characterised by a smaller symmetry group.
This is the case, for example, of the cubic-to-tetragonal structural phase transitions of anisotropically stressed perovskites like SrTiO$_3$, in which the $S_4\times \mathbb{Z}_2$ cubic symmetry  for $m>m_c$ is reduced to the $D_4\times\mathbb{Z}_2$ tetragonal one for $m<m_c$ \cite{BruceAhrony}.
The change to the reduced-symmetry form occurs in the vicinity of a crossover ``temperature'' $\tilde{m}= m_c(g) + \Delta \tilde{m}$ and can be characterized by a crossover exponent $\Phi$ according to $\Delta \tilde{m} \sim  g^{1/\Phi}$ \cite{Pfeuty}.
In the case of the cubic symmetry, we can introduce the following quadratic spin operator
\begin{equation}\label{eq:quadratictensor}
  \mathcal{E}_{i_1 i_2} = g \sum_{\langle \alpha, \beta \rangle} \left( \sigma^\alpha_{i_1} \sigma^\beta_{i_2} - \delta_{i_1i_2}\frac{\bm{\sigma}^2}{N} \right) \,,
\end{equation}
where again $\sigma^\alpha_i$ is the $i$-th component of a spin vector $\bm{\sigma}$ at the lattice site $\alpha$. We then expect two crossover exponents: one $\Phi_{\text{axis}}$ which characterizes the crossover to a phase where an axial interaction ($i_1=i_2$) is favoured and another one $\Phi_{\text{diag}}$ which instead favours the alignment along the diagonal ($i_1 \neq i_2$) of the cube.
In both cases we expect a crossover from an $N$-component to an Ising-like behavior, however, if the $N$-component transition is isotropic, then the two crossovers are equivalent since spins can be rotated so that $\mathcal{E}_{i_1 i_2}$ becomes $  \mathcal{E}_{i_1 i_1}$; if instead the transition is characterized by a truly cubic fixed point (this is the case of the standard cubic transition in $d=3$ for which we recall $N_c \simeq 2.9$), then the two crossover exponent becomes distinguished.  In fact, the separation of these two exponents can be used as an alternative criterion to determine $N_c$,
the other being the value of $N$ at which the $O(N)$ and hypercubic critical points coincide \cite{Carmona:1999}.

On the field-theoretical side, the discussion of quadratic operators is naturally encoded in the representation theory of the hypercubic group (a detailed analysis oriented to the construction of the spectrum of composite operators can be found in Ref.~\cite{Antipin:2019}). Here we sketch the essentials needed for this paper. For this purpose, it is useful to consider the following expression for the fundamental $N$-dimensional representation of $H_N$ (i.e.\ the lower dimension non-trivial representation)
\begin{equation}
    \phi_i = \Big(\,{\small{\Yvcentermath1\yng(2)\dots \yng(1)\,,\,\young(i)\,\Big)}} =(N-1,1)\,
\end{equation}
where the double partition $(\alpha,\beta)$ represents $\alpha$ objects, \textit{even} under $\mathbb{Z}_2$, which transform under an irreducible representation of $S_{\alpha}$, while the right partition  represents $\beta$ objects, \textit{odd} under $\mathbb{Z}_2$, which transform under the trivial representation of $S_\beta$. One can then construct the tensor product $\phi_i\otimes\phi_k$, which can be decomposed as follows
\begin{equation}\label{eq:tensorprodirreps}
\begin{split}
    \phi_{i}\otimes\phi_k & = \Big(\,{\small{\Yvcentermath1\yng(2)\dots \yng(1)\,,\,\young(i)\,}}\Big) \otimes {\small{\Big(\,\Yvcentermath1\yng(2)\dots \yng(1)\,,\,\young(k)\,}}\Big) \\
    & = \Big(\,{\small{\Yvcentermath1\yng(2)\dots \,,\, \emptyset \,}}\Big) \oplus \Big(\,{\small{\Yvcentermath1\yng(2,1)\dots \,,\, \emptyset \,}}\Big) \oplus \Big(\,{\small{\Yvcentermath1\yng(2)\dots \,,\, \yng(2) \,}}\Big)
     \oplus \Big(\,{\small{\Yvcentermath1\yng(2)\dots \,,\, \yng(1,1) \,}}\Big)
\end{split}
\end{equation}
and we start filling the tableau with indices that are all different. Each Young tableau then corresponds to an irrep of $H_N$. Since the left partition represents objects which are even under $\mathbb{Z}_2$, we associate an indexed field raised to the zeroth power to the first row and a second field power to the second row.
We then symmetrize w.r.t.\ the rows and antisymmetrize w.r.t.\ the columns. The same holds for the $\beta$ partition with increasing odd powers of the field.
In terms of these simple rules, we are able to explicitly write the irreps considered as
\begin{align}
    & S &&{\rm Singlet}  && \Big(\,{\small{\Yvcentermath1\yng(2)\dots \,,\, \emptyset \,}}\Big)  = \phi^2 &  \label{eq:singlet}\\
    & X &&{\rm diagonal~symmetric} &&\Big(\,{\small{\Yvcentermath1\yng(2,1)\dots \,,\, \emptyset \,}}\Big)  = \phi_i^2 - \phi_k^2 & \label{eq:symmetric}\\
    & Y &&{\rm off-diagonal~symmetric} && \Big(\,{\small{\Yvcentermath1\yng(2)\dots \,,\, \yng(2) \,}}\Big)  = \phi_i \phi_k & \label{eq:antisymmetric}
\end{align}
and the last irrep in Eq.~\eqref{eq:tensorprodirreps} is anti-symmetric, thus it cannot appear in quadratic oeprators with no derivatives and,
consequently, it will be discarded in the following.
We understand therefore that the tensor product in Eq.~\eqref{eq:tensorprodirreps} is decomposed into the direct sum of the trivial representation $S$, a traceless diagonal symmetric irrep $X$, an off-diagonal symmetric $Y$ and finally an anti-symmetric one $A$.\footnote{The corresponding $H_N$ tensor cannot be built with only two fields.} Since these irreducible representations are the representations carried by quadratic \emph{scaling} operators, we can eventually assign an anomalous dimension $\gamma_R$ for $R=S,X,Y$ to each one of them.
For this purpose, one can introduce the following four projectors in the space of quadratic operators
\begin{equation}
    \begin{split}
          P^{(S)}_{i_1 i_2 i_3 i_4} & = \frac{1}{N} \, \delta_{i_1 i_2} \delta_{i_3 i_4} \,,  \\
          P^{(X)}_{i_1 i_2 i_3 i_4} & = \delta_{i_1 i_2 i_3 i_4} - \frac{1}{N} \delta_{i_1 i_2} \delta_{i_3 i_4} \,, \\
          P^{(Y)}_{i_1 i_2 i_3 i_4} & = -\delta_{i_1 i_2 i_3 i_4} + \frac{1}{2} \left( \delta_{i_1 i_3} \delta_{i_2 i_4} +  \delta_{i_1 i_4} \delta_{i_2 i_3} \right) \,,
    \end{split}
\end{equation}
in terms of which we can decompose the general quadratic operator with no derivatives as
\begin{equation}\label{eq:quadratic-decomposition}
    \phi_{i_1}\phi_{i_2} =
    P^{(S)}_{i_1i_2i_3i_4}\phi_{i_3}\phi_{i_4} +
    P^{(X)}_{i_1i_2i_3i_4}\phi_{i_3}\phi_{i_4} + P^{(Y)}_{i_1i_2i_3i_4}\phi_{i_3}\phi_{i_4}\,.
\end{equation}
It could be checked that these projectors satisfy the following relations
\begin{equation}
\begin{split}
    P^{(R)}_{i_1 i_2 k l} P^{(R)}_{k l i_3 i_4} & = P^{(R)}_{i_l i_2 i_3 i_4} \,,\\
    P^{(R)}_{i_1 i_2 i_3 i_4} \delta_{i_1 i_4}\delta_{i_2 i_3} & = \dim(R)\,,
\end{split}
\end{equation}
where $\dim(R)$ stands for the dimension of the representation $R$. In the case at hand we have that $\dim(S)=1$, $\dim(X)=N-1$ and $\dim(Y)=N(N-1)/2$. We conclude by stating that the quadratic  deformation $X$ can be related to the diagonal crossover exponent $\Phi_{\text{diag}}$, while the deformation $Y$ to the axial one $\Phi_{\text{axis}}$.

\section{Effective potentials and their renormalization}
\label{sec:fprg}

The decomposition in subalgebras carried out in Section~\ref{sec:hypercubicalgebra} can be rephrased as follows: each of the $\mathcal{R}^{H_N}_{2n}$ can be labelled with the critical dimension at which the corresponding polynomials are canonically dimensionless, namely $d_{c} =  \frac{2n}{n-1}$. In the functional perturbative RG, on the other hand, the upper critical dimension $d_{c}$ (and therefore the label $n$) is taken as an input for the classification of multicritical universality classes \cite{ODwyer:2007, Codello:2017a}. The typical starting point to analyse the critical behavior of a given theory is the following GL action
\begin{equation}\label{eq:GLaction}
  S[\phi] = \int_x  \left\{\frac{1}{2} (\nabla \phi)^2  + V(\phi) \right\} \,.
\end{equation}
With the only input of $d_{c}$, one can derive very general expressions for the renormalization group flows $\beta_V$ of the effective potential $V(\phi)$
and $\beta_{Z_{ij}}$ of a general wavefunction, which is related to the anomalous dimension matrix $\gamma_{ij}=-\frac{1}{2}Z^{-1}_{ik} \beta_{Z_{kj}}$
(the latter will be given in the form of the singlet $\gamma_{ij}\phi_i\phi_j$).
In the following, we display these beta functionals using a diagrammatic representation in which $k$-vertices correspond to $k$ derivatives
of the potential w.r.t.\ the fields, namely
\begin{equation}
 \begin{split}
   V_{i_1 \,\cdots \, i_k} \equiv \frac{\partial^k V}{\partial\phi_{i_1} \cdots \partial \phi_{i_k}}
  = \begin{tikzpicture}[baseline=-.1cm]
      \draw (1,.5) to[out=180,in=90] (0,-.5);
      \draw (1,.5) to[out=180+20,in=90] (.4,-.5);
      \draw (1,.5) to[out=180+52,in=90] (.8,-.5);
      \draw (1,.5) to[out=0,in=90] (2,-.5);
      \filldraw [gray!50] (1,.5) circle (2pt);
      \draw(1,.5) circle (2pt);
      \filldraw [white!50] (0,-.5) circle (2pt);
      \draw(0,-.5) circle (2pt);
      \filldraw [white!50] (.4,-.5) circle (2pt);
      \draw(.4,-.5) circle (2pt);
      \filldraw [white!50] (.8,-.5) circle (2pt);
      \draw(.8,-.5) circle (2pt);
      \filldraw [white!50] (2,-.5) circle (2pt);
      \draw(2,-.5) circle (2pt);
       \draw[] (0,-.5) node[below] { ${\scriptstyle i_1}$};
       \draw[] (.4,-.5) node[below] { ${\scriptstyle i_2}$};
       \draw[] (.8,-.5) node[below] { ${\scriptstyle i_3}$};
       \draw[] (2,-.5) node[below] { ${\scriptstyle i_k}$};
       \draw[] (1.4,-.5) node[below] { ${\scriptstyle \cdots}$};
      \end{tikzpicture}
  \,,
 \end{split}
\end{equation}
in which the gray circle represents the $j$-vertex in question, hollow circles are the other vertices, and lines organize the contraction of indices. For example, the following diagram represents
\begin{equation}
\begin{split}
        \begin{tikzpicture}[baseline=-.1cm]
        \draw (1.5,0) circle (.5cm);
        \draw (1,0) to[out=50,in=130] (2,0);
        \draw (1,0) to[out=0,in=180] (2,0);
        \draw (1,0) to[out=-50,in=-130] (2,0);
        \filldraw [gray!50] (1.5,.5) circle (2pt);
        \draw(1.5,.5) circle (2pt);
        \filldraw [gray!50] (1,0) circle (2pt);
        \draw (1,0) circle (2pt);
        \filldraw [gray!50] (2,0) circle (2pt);
        \draw (2,0) circle (2pt);
      \end{tikzpicture}
& = V_{i_1 i_2}V_{i_1 i_3 i_4 i_5 i_6}V_{i_2 i_3 i_4 i_5 i_6} \,.
\end{split}
\end{equation}

It is worth sidestepping for a moment and consider one of the most remarkable and striking consequences of the functional formalism, namely the fact that some of the well-known expressions derived for the beta functionals in the single-component field case \cite{ODwyer:2007} can be directly
enhanced to their multicomponent counterparts without performing any single additional computation \cite{Zinati:2019gct}. Few remarks are in order.
The basic idea is that the multicomponent ($N > 1$) beta functionals $\beta_V$ and $\beta_Z$
must reduce to their known single-component cases in the $N = 1$ limit. In the single-component case, typical monomials contributing to  $\beta_V$ and {\color{black}$\beta_Z$} are constructed in terms of derivatives $V^{(k)}$ of the potential, possibly linked by the appropriate powers of $V^{(2)}$ insertions to have proper vacuum diagrams. In promoting these monomials to the multicomponent case, namely rewriting $V^{(2)} \to V_{i_1 i_2}$ and $V^{(k)} \to V_{i_1 \dots i_k}$, we notice that in dimensional regularization, self-contractions like $V_{i_1i_1 \dots i_k}$ or traced mass insertions like $V_{i_1i_1}$ (which are generated by the expansion of the propagators in the loop diagrams) do not contribute. Therefore the relative coefficients appearing in front of the monomials in the single-component case  are univocally fixed and stay the same in the $N > 1$ case.

The single-component functional RG flows for the potential and the wavefunction can be found in \cite{ODwyer:2007, Codello:2017a, Codello:2017epp}
and can be used to obtain their generalizations.
Let us emphasize that the generalizations to the multicomponent cases of the RG equations for $d_c=6,4,3$ have already appeared in \cite{Osborn:2017ucf}, but here we want to make the general point that we can generalize the RG equations for any $d_{c,m}$ to some extent.

For \emph{even} potentials (apart the trivial LO $d_c = 4$ case), in the NLO beta functional
$\beta_V^{\text{NLO}}$, there appears only one term involving $V_{i_1i_2}$ and this term is linear in the mass insertion \cite{ODwyer:2007}. Because of this fact, there is only one multicomponent diagram generalizing this term, and its relative coefficient can be fixed univocally. A useful example is the $d_c = 3$ case \cite{ODwyer:2007}: the $N = 1$ beta functional is given
by
\begin{equation}\label{eq:exampled3}
\beta_{V} = \frac{1}{3} (V^{(3)})^2 + \frac{1}{6} V^{(2)}(V^{(5)})^2 - \frac{4}{3} V^{(3)}V^{(4)}V^{(5)} - \frac{\pi^{2}}{12}(V^{(4)})^3 \,,
\end{equation}
for which the NLO monomial $V^{(2)}(V^{(5)})^2$ is the only involving a mass insertion and the $V_{i_1 i_2}$ can only connect the two five vertices. The vertices in other NLO monomials can be connected only following the single component loop diagram from which they come from.
Therefore, the generalization to \eqref{eq:exampled3} reads
\begin{equation}
\begin{split}
\beta_{V}	& = \frac{1}{3} V_{i_1i_2i_3}V_{i_1i_2i_3} + \frac{1}{6} V_{i_1i_2}V_{i_1i_3i_4i_5i_6}V_{i_2i_3i_4i_5i_6}\\
& -\frac{4}{3} V_{i_1i_2i_3}V_{i_3i_4i_5i_6}V_{i_1i_2i_4i_5i_6}-\frac{\pi^{2}}{12}V_{i_1i_2i_3i_4}V_{i_3i_4i_5i_6}V_{i_1i_2i_5i_6} \,.
\end{split}
\end{equation}

In the {\textit{odd}} case it is possible to derive a general formula for the $N=1$ beta functional $\beta_V$ only at LO (see Ref.~\cite{Codello:2017a}). Since this formula can be obtained by analytical continuation of the NLO {\textit{even}} case, it involves only one $V_{i_1 i_2}$ in only one monomial and, as before, its multicomponent counterpart can be fixed univocally. The picture is different for the {\textit{odd}} NLO case for which,
apart the $d_c=6$ case, no general functionals $\beta_V$ and $\beta_Z$ are known. In the $N=1$ single component $d_c=6$ case, the beta functional is $\beta_V = a (V^{(2)})^3 + b (V^{(2)})^2 (V^{(3)})^3$ and since $V^{(2)}$ are mass insertions and since the NLO is at two loop (vacuum sunset diagram), there are three propagators where the three mass insertions can fit
\begin{align}
(V^{(2)})^3(V^{(3)})^2  \to & ~\alpha ~ V_{i_1i_2}V_{i_2i_3}V_{i_3i_4}V_{i_1i_5i_6}V_{i_4i_5i_6}  \nonumber\\ + & ~\beta  ~ V_{i_1i_2}V_{i_2i_3}V_{i_4i_5}V_{i_1i_4i_6}V_{i_3i_5i_6} \nonumber\\
+ & ~\gamma ~ V_{i_1 i_2}V_{i_3i_4}V_{i_5i_6}V_{i_1i_3i_5}V_{i_2i_4i_6}\,,
\end{align}
and thus the knowledge of the $N=1$ case only fixes the sum $\alpha+\beta+\gamma$ \cite{Osborn:2017ucf}. Finally, in the case of the LO ({\textit{even}} and {\textit{odd)}} $\beta_Z$ there is only one monomial with no $V_{i_1 i_2}$ mass insertions and the results of the {\textit{even}} case agrees with CFT results \cite{Codello:2017qek}. The determination of the NLO contributions in the {\textit{odd}} case will complete the knowledge of the universal data of one component scalar field theories.
The discussion above provides the rationale behind the analysis of the critical content of scalar field theories endowed with hypercubic symmetry using the functional perturbative RG.
In this case, taking advantage of the results of Section~\ref{sec:hypercubicalgebra}, the GL effective potential of Eq.~\eqref{eq:GLaction} can be expressed as
\begin{equation}
  V(\phi) = \sum_n \sum_{j=1}^{p(n)} \lambda_{j,n} \, I_j^{(2n)} \,,
\end{equation}
namely, we assign a coupling constant $\lambda_{j,n}$ to each of the $p(n)$ invariant polynomials $I^{(2n)}_j$ composing the corresponding $\mathcal{R}^{H_N}_{2n}$ subspace. For a given $d_{c}$, dimensionful $\beta$-functions for the couplings $\lambda_{j,n}$ follow by inserting the explicit form of the potential in the corresponding $\beta_V$ and by simplifying the algebra of Kronecker symbols resulting from contracting the fields. In terms of the fixed points of the associated system of dimensionless beta functions, one can finally express universal quantities, such as critical exponents, as power series in $\epsilon = d_{c} - d$.

\subsection{Universal quantities}

In dealing with the quadratic operators of Section~\ref{sec:quadraticdeformations}, we can use Eq.~\eqref{eq:quadratic-decomposition} to source individually each irrep contribution in the path integral and to renormalize them as composite operators. We introduce three sources $J^{(R)}$, for $R=S,X,Y$, in the combination
\begin{equation} \label{eq:quadratic-composite}
    {\cal O}_{\rm qu}(\phi) = \sum_{R=S,X,Y} J^{(R)}_{i_1 i_2} P^{(R)}_{i_1i_2i_3i_4}\phi_{i_3}\phi_{i_4}\,.
%    J^{(S)}_{i_1 i_2} P^{(S)}_{i_1i_2i_3i_4}\phi_{i_3}\phi_{i_4} +
%    J^{(X)}_{i_1 i_2} P^{(X)}_{i_1i_2i_3i_4}\phi_{i_3}\phi_{i_4} +
%    J^{(Y)}_{i_1 i_2} P^{(Y)}_{i_1i_2i_3i_4}\phi_{i_3}\phi_{i_4}\,.
\end{equation}
Naturally, the indices of the source $J^{(S)}$ of the scalar $S$ operator are redundant as it couples to its operator only through the trace, but contractions are accounted for by the projectors.
The computation of the expectation value $\langle {\cal O}_{\rm qu}(\phi) \rangle$ requires renormalization of the three sources,
corresponding to the above three operators, and consequently gives three independent gamma functions.
For almost all values of $N$, the three operators are renormalized multiplicatively and do not mix, that is, they are already scaling operators
(scaling operators thus carry a scaling dimension and a label of the irreps of $H_N$).

From our point of view, the most convenient route for the computation
is to determine the renormalization group running of the three composite operators using
$\beta_V$. In fact, the advantage of using a functional formalism is that the running of the potential $V(\phi)$ already includes all the information on the running of all relevant operators, including \eqref{eq:quadratic-composite}.
To extract this information it is sufficient to replace $V(\phi) \to V(\phi)+{\cal O}_{\rm qu}(\phi)$ in $\beta_V$
and retain only the leading order in the three source couplings $J^{(R)}$. By construction, this operation provides the gamma functions $\gamma_R$ of the three operators multiplied by the corresponding sources, which can easily be related to either the critical exponents $\theta_R$ or the scaling dimensions $\Delta_R$.\footnote{Notice that for almost all values of $N$ the irreducible operators appearing in \eqref{eq:quadratic-composite} are automatically scaling operators, which means that there is no mixing induced by radiative corrections to our decomposition. In other words, the generator of the dilatations ${\cal D}$ commutes with the action of the hypercubic group on the irreducible decomposition. This is not true, however, for all values of $N$.
In fact, by changing appropriately $N$ we can end up in the situation in which two or more scaling operators degenerate to the same multiplet because their scaling dimensions coincide. This results in a nontrivial Jordan cell and induces a logarithmic contribution.
Interestingly, this happens for the special limits $N=0$ (see Ref.~\cite{Zinati:2020xcn}) and $N=-2$ of the hypercubic model, which is obviously reminiscent of the same limit of the $O(N)$ model that has recently been associated with the loop erased random walk \cite{Wiese1}.
}
In particular, the critical exponents $\theta_R$ are obtained as the negatives of eigenvalues of the stability matrix $M_{ij} = \partial_j \beta_i$.
In  bridging to CFT, we recall that the eigenvalues $\theta_R$ are related to the CFT scaling dimension $\Delta_R$ of scaling operators by the relation $\theta_R = d - \Delta_R$.
Alternatively, one can retrieve the gamma functions $\gamma_R$ as $\gamma_R = \theta_R - (d + 2 - \eta )/2$, where the  anomalous dimension critical exponent $\eta$ is obtained by diagonalizing the matrix $\gamma_{ij}$ with an additional factor $2$.
In particular, the correlation length critical exponent $\nu$ is related to the inverse of the quadratic singlet in $V(\phi)$ as $\nu = \theta_S^{-1}$, while the crossover exponents discussed in Section \ref{sec:quadraticdeformations} are found to be $\Phi_{\text{axis}} = \theta_X/\theta_S$ and $\Phi_{\text{diag}} = \theta_Y/\theta_S$.

Finally, we compute the $A$-function, which is the scalar function from which one can derive the RG flow as a gradient, for all the even models. This can be of particular interest as it entails very important implications on the asymptotic behavior of the renormalization group equations, as well as on the classification of renormalization-group fixed points \cite{Wallace:1974, Wallace:1974dy, Osborn:2017ucf}. If we call $\lambda_a$ all the couplings of interest and $\beta_a$ the corresponding $\beta$-functions, then $A$ is derived implicitly from
\begin{equation}
  \beta_a = \sum_b h_{ab} \frac{\partial A}{\partial \lambda_b} \,,
\end{equation}
and $h_{ab}$ is a suitable positive metric in the space of all couplings.
We require the metric to be flat in the space of marginal couplings parametrized as the symmetric tensors $V_{i_1,\cdots, i_{2n}}$,
which can be always be done at leading and next-to-leading order \cite{Wallace:1974, Wallace:1974dy}.
Consequently, we express the $A$-function from contractions of $V_{i_1,\cdots, i_{2n}}$ when possible,
and it is understood that beta functions are derived from the gradients and there is a flat metric.\footnote{%
An important reason to express $A$ using the general couplings $V_{i_1,\cdots, i_{2n}}$ instead of the specific
$H_N$ bases is that, using this definition, we can compare the values even with models with symmetry different from $H_N$.
}
%where in the rest of the paper we assume the metric $h_{ab}$ in the space of couplings to be flat.
%The $A$ function is normalized such that its gradient gives the flow of the marginal couplings when parametrized
%\begin{align}
%  \beta_{ijklmn} =\frac{\partial A}{\partial_{V_{ijklmn}}} & \qquad {\rm for } \quad d_c=3 \\
%  \beta_{ijklmnpq} =\frac{\partial A}{\partial_{V_{ijklmnpq}}} & \qquad {\rm for } \quad d_c=\frac{8}{3}\,.
%\end{align}

\section{Tricritical models in \texorpdfstring{\bm{$d = 3-\epsilon$}}{3-eps}}
\label{sec:3-eps}

The discussion of the critical properties in $d=3-\epsilon$ dimensions would require the simultaneous fine-tuning at criticality of other two parameters beside the mass, namely the two quartic non-linearities $\lambda_1$ and $\lambda_2$. This being said, in the following we shall still refer to RG fixed points possibly  emerging at $d_c=3$ as tricritical ones, despite their multicritical nature.
Critical properties can be accessed in terms of the following potential
\begin{equation}\label{eq:pot6}
\begin{split}
V(\phi)  & =  \sigma_1\, \Bigl( \sum_i \phi_i^2 \Bigr)^3  +	\sigma_2\, \sum_i \phi_i^2 \, \sum_j \phi_j^4 + \sigma_3\, \sum_i \phi_i^6 \\
& + \lambda_1\, \Bigl( \sum_i \phi_i^2 \Bigr)^2 + \lambda_2\, \sum_i \phi_i^4 + m^2\, \phi_i^2 \,.
\end{split}
\end{equation}
In this case, LO and NLO beta functionals $\beta_V$ and $\beta_Z$ and the expression for the $A$-function are given by
\begin{equation}
\begin{split}
        \beta_V  = &
        + \frac{1}{3} ~
        \begin{tikzpicture}[baseline=-.1cm]
        \draw (0,0) circle (.5cm);
        \draw (-.5,0) to [out=0,in=180] (.5,0);
        \filldraw [gray!50] (.5,0) circle (2pt);
        \draw (.5,0) circle (2pt);
        \filldraw [gray!50] (-.5,0) circle (2pt);
        \draw (-.5,0) circle (2pt);
        \end{tikzpicture}
        ~ +\frac{1}{6} ~
        \begin{tikzpicture}[baseline=-.1cm]
        \draw (1.5,0) circle (.5cm);
        \draw (1,0) to[out=50,in=130] (2,0);
        \draw (1,0) to[out=0,in=180] (2,0);
        \draw (1,0) to[out=-50,in=-130] (2,0);
        \filldraw [gray!50] (1.5,.5) circle (2pt);
        \draw(1.5,.5) circle (2pt);
        \filldraw [gray!50] (1,0) circle (2pt);
        \draw (1,0) circle (2pt);
        \filldraw [gray!50] (2,0) circle (2pt);
        \draw (2,0) circle (2pt);
        \end{tikzpicture}
        ~ - \frac{4}{3} ~
        \begin{tikzpicture}[baseline=-.1cm]
        \draw (3,0) circle (.5cm);
        \draw (2.531,-.171) to [out=-30,in=-150] (3.469,-.171);
        \draw (2.531,-.171) to [out=30,in=150] (3.469,-.171);
        \draw (3,.5) to[out=-90,in=150] (3.469,-.171);
        \filldraw [gray!50] (3,.5) circle (2pt);
        \draw (3,.5) circle (2pt);
        \filldraw [gray!50] (3.469,-.171) circle (2pt);
        \draw (3.469,-.171) circle (2pt);
        \filldraw [gray!50] (2.531,-.171) circle (2pt);
        \draw (2.531,-.171) circle (2pt);
        \end{tikzpicture}
        ~ - \frac{\pi}{12} ~
        \begin{tikzpicture}[baseline=-.1cm]
        \draw (4.5,0) circle (.5cm);
        \draw (4.031,-.171) to [out=-30,in=-150] (4.969,-.171);
        \draw (4.5,.5) to [out=-90,in=30] (4.031,-.171);
        \draw (4.5,.5) to[out=-90,in=150] (4.969,-.171);
        \filldraw [gray!50] (4.5,.5) circle (2pt);
        \draw (4.5,.5) circle (2pt);
        \filldraw [gray!50] (4.969,-.171) circle (2pt);
        \draw (4.969,-.171) circle (2pt);
        \filldraw [gray!50] (4.031,-.171) circle (2pt);
        \draw (4.031,-.171) circle (2pt);
     \end{tikzpicture} ~, \\
%
%     \beta_Z  = &
      \gamma_{ij}\phi_i\phi_j = &
     + \frac{1}{90} ~
     \begin{tikzpicture}[baseline=-.1cm]
     \draw (1.5,0) circle (.5cm);
     \draw (1,0) to[out=50,in=130] (2,0);
     \draw (1,0) to[out=0,in=180] (2,0);
     \draw (1,0) to[out=-50,in=-130] (2,0);
     \filldraw [gray!50] (1,0) circle (2pt);
     \draw (1,0) circle (2pt);
     \filldraw [gray!50] (2,0) circle (2pt);
     \draw (2,0) circle (2pt);
     \end{tikzpicture} ~,\\
     A = & + \epsilon ~
     \begin{tikzpicture}[baseline=-.1cm]
     \draw (0,0) circle (.5cm);
     \draw (-.5,0) to[out=30,in=150] (.5,0);
     \draw (-.5,0) to[out=-30,in=-160] (.5,0);
     \draw (-.5,0) to[out=60,in=180] (0,.32);
     \draw (0,.32) to[out=0,in=130] (.5,0);
     \draw (-.5,0) to[out=-60,in=180] (0,-.32);
     \draw (0,-.32) to[out=0,in=-130] (.5,0);
     \filldraw [green!50] (-.5,0) circle (2pt);
     \draw (-.5,0) circle (2pt);
     \filldraw [green!50] (.5,0) circle (2pt);
     \draw (.5,0) circle (2pt);
    \end{tikzpicture}
    ~ + \frac{20}{9} ~
    \begin{tikzpicture}[baseline=-.1cm]
    \draw (1.5,0) circle (.5cm);
    \draw (1.031,-.171) to [out=-30,in=-150] (1.969,-.171);
    \draw (1.031,-.171) to [out=35,in=250] (1.5,.5);
    \draw (1.031,-.171) to [out=80,in=210] (1.5,.5);
    \draw (1.969,-.171)  to [out=100,in=-30] (1.5,.5);
    \draw (1.969,-.171)  to [out=145,in=290] (1.5,.5);
    \draw (1.031,-.171)  to [out=30,in=150]  (1.969,-.171);
    \filldraw [green!50] (1.5,.5) circle (2pt);
    \draw (1.5,.5) circle (2pt);
    \filldraw [green!50] (1.969,-.171) circle (2pt);
    \draw (1.969,-.171) circle (2pt);
    \filldraw [green!50] (1.031,-.171) circle (2pt);
    \draw (1.031,-.171) circle (2pt);
    \end{tikzpicture} ~.
     \end{split}
\end{equation}

The $\beta$-functions of the couplings, the anomalous dimension matrix $\gamma_{ij}$ and the $A$ function are then readily obtained by inserting the effective potential of Eq.~\eqref{eq:pot6} in the $\beta$-functionals above and by simplying the algebra of Kronecker symbols that are responsible for the field contractions in $V(\phi)$. The $\beta$-function of the quadratic (mass) coupling is
\begin{align}
  \beta_{m^2}
  &
  = - 2 m^2 + \frac{2}{9} \left[(N+2)\lambda_1^2 +6 \lambda_1 \lambda_2+3 \lambda_2^2\right] \nonumber\\
  &
  + \frac{16}{675} \left[\left(N^2\!+\!6 N\!+\!8\right)\sigma_1^2  + 6 (N\!+\!4)\sigma_1 \sigma_2  + 30 \sigma_1 \sigma_3) +  (N\!+\!14) \sigma_2^2 + 30 \sigma_2 \sigma_3 + 15 \sigma_3^2\right] m^2 \nonumber\\
  &
  -\frac{1}{540} \left[\pi ^2 (N+8)+96\right] \left[\left(N^2+6 N+8\right)\sigma_1 +3
   (N+4)\sigma_2 + 15 \sigma_3 \right] \lambda_1^2 \nonumber\\
  & - \frac{1}{60} \left\{\left[3 \pi ^2 (N+4)+160\right]\sigma_1  + \left[\pi ^2
   (N+14)+160\right] \sigma_2 + 5 \left(32+3 \pi ^2\right) \sigma_3\right\} \lambda_2^2 \nonumber\\
  &
  -\frac{1}{90} \left[\pi ^2 N^2+12 \left(8+\pi ^2\right) N+32 \left(12+\pi
   ^2\right)\right] \lambda_1\lambda_2\sigma_1 -\frac{1}{6} \left(32+3 \pi
    ^2\right) \lambda_1\lambda_2\sigma_3\nonumber\\
  &
  -\frac{1}{90} \left[\left(5 \pi ^2 +32\right) N+40 \pi ^2+448\right] \lambda_1\lambda_2\sigma_1 \,.
\end{align}
The $\beta$-functions for quartic couplings are given by
\begin{align}
  \beta_{\lambda_1} & = - (1 + \epsilon) \lambda_1  + \frac{8}{15} \left[(N+4)\lambda_1 \sigma_1  + 2 \lambda_1 \sigma_2 + 3\sigma_1\lambda_2  + \sigma_2\lambda_2 \right] \nonumber\\
   & -\frac{1}{1350} (N+4) \left[3 \pi ^2 N^2+\left(680+54 \pi ^2\right) N+348 \pi ^2+4528\right] \lambda_1\sigma_1^2 + \frac{32}{45} \lambda_1\sigma_3^2 \nonumber \\
   &
   - \frac{1}{225}\left[3 \pi ^2 N^2 + 2 \left(292+27 \pi ^2\right) N + 8 \left(508+39 \pi^2\right)\right]  \lambda_1\sigma_1 \sigma_2
   -\frac{4}{45} \left(22+3 \pi ^2\right) \lambda_1 \sigma_2 \sigma_3
   \nonumber\\
   &
   - \frac{1}{675} \left[\left(92+15 \pi ^2\right) N + 246 \pi ^2 + 2872 \right] \lambda_1\sigma_2^2
   - \frac{1}{45} \left[3 \pi ^2 (N+10) + 392\right] \lambda_1\sigma_1 \sigma_3 \nonumber\\
   & -\frac{1}{150} \left[\pi ^2 N^2\!+\!20 \left(12\!+\!\pi ^2\right)\! N + 8 \left(216+17 \pi^2\right)\right]\lambda_2\sigma_1^2 - \frac{1}{450} \left[3 \pi ^2(N\!+38)\!+\!1376\right] \lambda_2\sigma_2^2  \nonumber\\
   &
   -\frac{1}{75} \left[(5 \pi ^2 \!+32) N \!- 80 \pi ^2\!-\!992\right] \lambda_2\sigma_1\sigma_2
   - \frac{1}{5} \left(3 \pi ^2 \!+ 32\right)\lambda_2\sigma_1\sigma_3
   -\frac{1}{15} \left(32 \!+ 3 \pi ^2\right) \lambda_2 \sigma_2 \sigma_3 \,,
\end{align}
and
\begin{align}
  \beta_{\lambda_2} & = - (1 + \epsilon) \lambda_2 + \frac{8}{45} \left[\lambda_1 \sigma_2 (N+8)+15 \lambda_1 \sigma_3 +6 \lambda_2
   \sigma_1 + 12 \lambda_2\sigma_2 + 15 \lambda_2\sigma_3 \right]\nonumber\\
    &
    -\frac{2}{675} \left[-10 N^2 + (9 \pi ^2 +156) N + 144 \pi ^2+1936\right] \lambda_2
    \sigma_1^2 - \frac{1}{90} \left(1736 + 135 \pi ^2\right) \lambda_2
    \sigma_3^2 \nonumber\\
    &
    - \frac{1}{675} \left[(9 \pi ^2 + 148) N + 828 \pi ^2 + 10808\right] \lambda_2
    \sigma_2^2 -\frac{2}{45} \left(820+63 \pi ^2\right) \lambda_2 \sigma_2\sigma_3 \nonumber\\
    &
    -\frac{4}{225} \left[(3 \pi ^2 +50) N+102 \pi ^2+1352\right] \lambda_2 \sigma_1 \sigma_2 -\frac{4}{45} \left(362+27 \pi ^2\right) \lambda_2 \sigma_1 \sigma_3 \nonumber \\
    &
    -\frac{1}{450} \left[\pi ^2 N^2 + 4 \left(124+9 \pi ^2\right) N + 16 \left(388+29 \pi
   ^2\right)\right] \lambda_1 \sigma_2^2 -\frac{1}{2} \left(40 + 3 \pi ^2\right) \lambda_1\sigma_3^2 \nonumber \\
   &
   -\frac{2}{225} (N+8) \left[\left(16+\pi ^2\right) N +16 \left(13+\pi ^2\right)\right] \lambda_1 \sigma_1 \sigma_2 \nonumber\\
   &
   -\frac{2}{15} \left[\left(16+\pi ^2\right) N + 16 \left(13+\pi ^2\right)\right]\lambda_1 \sigma_1 \sigma_3 \nonumber\\
   &
   - \frac{1}{45} \left[\left(32+3 \pi ^2\right) N + 8 \left(202+15 \pi ^2\right)\right] \lambda_1 \sigma_2 \sigma_3 \,.
\end{align}
The RG of the marginal (sextic) couplings read
\begin{align}
  \beta_{\sigma_1} & = - 2 \sigma_1 \epsilon + \frac{4}{15} \left[\sigma_1^2 (3 N + 22) + 12 \sigma_1 \sigma_2 + \sigma_2^2 \right] - \frac{1}{450} \left[\pi ^2 N^3 + 2 \left(212 + 17 \pi ^2 \right) N^2 \right. \nonumber\\
  & \left. +4\left(1716+155 \pi ^2\right) N + 32 \left(826+85 \pi ^2\right)\right] \sigma_1^3 -\frac{1}{30} \left[3 \pi ^2 (N+16)+560\right] \sigma_1^2 \sigma_3 \nonumber\\
  &  - \frac{1}{50} \left[\pi ^2 N^2+4 \left(88+7 \pi ^2\right) N + 112
   \left(32+3 \pi ^2\right)\right] \sigma_1^2 \sigma_2 + \frac{16}{15} \sigma_1\sigma_3^2 - \frac{4}{5} \left(10+\pi ^2\right) \sigma_1 \sigma_2 \sigma_3 \nonumber\\
   & -\frac{1}{75} \sigma_1\sigma_2^2 \left[ \left(36 + 5 \pi ^2\right)N+142 \pi ^2+1736 \right]  - \frac{1}{30}
   \left(32+3 \pi ^2\right) \sigma_2^2 \sigma_3 \nonumber\\
   & -\frac{1}{450}\left[(N + 74)\pi ^2  + 960\right] \sigma_2^3 \,,
\end{align}
and
\begin{align}
  \beta_{\sigma_2} & = - 2 \sigma_2 \epsilon  + \frac{4}{15} \left[ 13 \sigma_2^2 + (2 N + 36) \sigma_2\sigma_1 + 10 \sigma_2\sigma_3 + 30 \sigma_1 \sigma_3 \right] \nonumber\\
  & -\frac{2}{225} \left[\left(26+3 \pi ^2\right) N + 291 \pi ^2 + 3124\right] \sigma_2^3  - \frac{1}{30} \left(1376 + 117 \pi ^2\right) \sigma_2^2 \sigma_3 \nonumber\\
  & -\frac{1}{150} \left[\pi ^2 N^2 + 8 \left(112+9 \pi ^2\right) N + 8 \left(2424 + 241 \pi^2 \right)\right] \sigma_2^2 \sigma_1 -\frac{1}{30} \left(568 + 45 \pi ^2\right) \sigma_2 \sigma_3^2 \nonumber \\
  & -\frac{2}{75} \left[\left(18+\pi ^2\right) N^2+\left(644+57 \pi ^2\right) N + 512 \pi
   ^2+4816\right] \sigma_2 \sigma_1^2 -\frac{3}{2} \left(40+3 \pi ^2\right) \sigma_1 \sigma_3^2\nonumber\\
  & - \frac{1}{15} \left[\left(32 \!+\! 3 \pi ^2\right) N + 6 \left(476 \!+\! 41 \pi ^2\right)\right] \sigma_1 \sigma_2 \sigma_3  -\frac{2}{15} \left[(3 \pi ^2 \!+\! 62)N + 102 \pi ^2+1064\right] \sigma_1^2\sigma_3 \,,
\end{align}
and finally
\begin{align}
  \beta_{\sigma_3} & = - 2 \sigma_3 \epsilon  + \frac{4}{45} \left[75 \sigma_3^2 + 60 \sigma_3 \sigma_1 + 120\sigma_2 \sigma_3 + \sigma_2^2 (N + 32) \right] -\frac{1}{30} \left(2248 + 225 \pi ^2\right)\sigma_3^3\nonumber\\
  & -\frac{2}{15} \left(1244+135 \pi ^2\right)\sigma_3 ^2 \sigma_1 -\frac{1}{15} \left(3088 + 315 \pi ^2\right) \sigma_3^2 \sigma_2  -\frac{2}{75} \left[(15 \pi ^2 + 204) N \right. \nonumber\\
  &\left. + 1020 \pi^2 + 9136\right] \sigma_1\sigma_2\sigma_3 - \frac{1}{450} \left[45 \pi ^2(N + 568) + 8280 \pi ^2 + 79472\right] \sigma_3\sigma_2^2  \nonumber\\
    & +\frac{2}{225} \left[8 N^2 - \left(45 \pi ^2 + 372\right) N - 720 \pi ^2 - 5936\right] \sigma_3 \sigma_1^2 \nonumber\\
  & -\frac{4}{225} \left[4 N^2 + \left(21 \pi ^2 + 248\right) N + 384 \pi ^2+3456\right] \sigma_1 \sigma_2^2 \,.
\end{align}
The general expression for the anomalous dimension $\eta$ can be derived from diagonalizing $\gamma_{ij}$ and it reads
\begin{equation}
\eta = \frac{1}{675} (N+2)(N+4) \sigma_1^2 + \frac{1}{675} (N+14)\sigma_2^2  + \frac{1}{45} \sigma_3^2 +  \frac{2}{225}(N+4)\sigma_1 \sigma_2 + \frac{2}{45} \sigma_1 \sigma_3 + \frac{2}{45} \sigma_2 \sigma_3 \,.
\end{equation}
while finally the expression for the $A$ functions is given by
\begin{align}
  A &
  = -\frac{1}{15} N \left[\left(N^2 \!+\! 6 N \!+\! 8\right)\sigma_1^2 + 6 (N\!+\!4)\sigma_1\sigma_2  + 30
   \sigma_1\sigma_3 + (N\!+\!14)\sigma_2^2 + 30 \sigma_2 \sigma_3+15 \sigma_3^2\right] \epsilon \nonumber\\
  &
  +\frac{4}{675} N \left(3 N^3+40 N^2+156 N+176\right) \sigma_1^3 +\frac{4}{75} N \left(3 N^2+34 N+88\right) \sigma_1^2\sigma_2 \nonumber\\
  &
  +\frac{4}{15} N (3 N+22) \sigma_1^2 \sigma_3 +\frac{4}{225} N \left(N^2+50 N+324\right) \sigma_1\sigma_2^2
  +\frac{8}{15}(N+24) \sigma_1\sigma_2 \sigma_3 + \frac{20}{3} \sigma_1\sigma_3^2 \nonumber\\
  & + \frac{4}{225} N (7 N+118) \sigma_2^3 +\frac{4}{45} N (N+74) \sigma_2^2\sigma_3 +\frac{20 }{3} N \sigma_2 \sigma_3^2 +\frac{20}{9} N \sigma_3^3 \,.
\end{align}
By taking derivatives of $A$ with respect the marginal couplings $\sigma_a$ it is straightforward to determine the metric $h_{ab}$
in the space of the couplings and check that it is positive.
We report also the anomalous dimensions associated to the quadratic deformations
\begin{align}\label{eq:quadratic3d}
  \gamma_S & =
  -\frac{1}{45} \sigma_1^2 (N\!+\!2) (N\!+\!4)
  -\!\frac{2}{15} \sigma_1 \sigma_2(N\!+\!4)
  -\!\frac{2}{3} \sigma_1 \sigma_3
  -\!\frac{1}{45} \sigma_2^2 (N\!+\!14)
  -\!\frac{2}{3} \sigma_2 \sigma_3
  -\!\frac{1}{3}  \sigma_3^2 ~,\\
  \gamma_X & =
  -\frac{1}{225} \sigma_1^2 (N\!+\!4) (N\!+\!10)
  -\!\frac{2}{225} \sigma_1 \sigma_2 (7 N\!+\!60)
  -\!\frac{2}{3}  \sigma_1 \sigma_3
  -\!\frac{7}{675} \sigma_2^2 (N\!+\!30)
  -\!\frac{2}{3} \sigma_2 \sigma_3
  -\!\frac{1}{3}  \sigma_3^2 ~,\\
  \gamma_Y & =
  -\frac{1}{225} \sigma_1^2 (N\!+\!4) (N\!+\!10)
  -\!\frac{2}{225} \sigma_1 \sigma_2(3N\!+\!28)
  -\!\frac{2}{15}  \sigma_1 \sigma_3
  -\!\frac{1}{675} \sigma_2^2 (N\!+\!42)
  -\!\frac{2}{45}  \sigma_2\sigma_3 ~.
\end{align}
It is straightforward to check that the gamma function of the singlet $S$ is related to the beta function of $m^2$
as $\gamma_S=2-\frac{\partial}{\partial m^2} \beta_{m^2}$, but, in general, $\beta_{m^2}$ contains more information.

{\makegapedcells
\begin{table}[h]
\begin{center}
\begin{tabular}{|  l | c | c | }
\hline
 & Tri-Ising & Tri-$O(N)$\\
 \hline
 $\eta$ &   $+\frac{1}{500}\epsilon^2$  & $+\frac{(N+2)(N+4)}{12(3N +22)^2} \epsilon^2$ \\
 $A$    &   $-\frac{3 N}{100}\epsilon^3$& $-\frac{5 N (N+2) (N+4)}{4 (3 N + 22)^2} \epsilon ^3$\\
 $\theta_S$ & $2-\frac{4}{125}\epsilon ^2$ & $2-\frac{4 \left(N^2+6 N+8\right)}{3 (3 N+22)^2} \epsilon ^2$\\
 $\theta_X$ & $2-\frac{4}{125}\epsilon ^2$ & $2-\frac{\left(N^2+12 N+32\right)}{3 (3 N+22)^2} \epsilon ^2$\\
 $\theta_Y$ & $2-\frac{1}{500}\epsilon ^2$ & $2-\frac{\left(N^2+12 N+32\right)}{3 (3 N+22)^2} \epsilon ^2$\\
\hline
\end{tabular}
\end{center}
\caption{Anomalous dimension $\eta$, $A$-function and critical exponents $\theta_R$ at the tricritical Ising and the tricritical $O(N)$ fixed points.}
\label{tab:tricriticalFPs}
\end{table}
}

The system of $\beta$-functions admits a total of $8$ fixed points (both real and complex). However, as a first physical requirement, we restrict ourselves to consider critical theories which are perturbatively unitary: we therefore select only real fixed points with positive anomalous dimension $\eta>0$ and characterized by a real spectrum.
A second physical requirement is that the fixed point potential \eqref{eq:pot6} is stable, namely, bounded from below: this amounts at considering fixed points for which $(\sigma^{\star}_1 + \sigma^{\star}_2 + \sigma^{\star}_3) > 0$.
Under these conditions, beside the trivial Gaussian fixed point, the system of $\beta$-functions predicts the tricritical decoupled Ising universality class  with coordinates $\{\sigma^{\star}_1 = 0 , \sigma^{\star}_2 = 0 , \sigma^{\star}_3 = \frac{3}{10} \epsilon  \}$ and the tricritical $O(N)$ universality class with coordinates
$\{\sigma^{\star}_1 = \frac{15 \epsilon}{2 (3N + 22)} , \sigma^{\star}_2 = 0 , \sigma^{\star}_3 = 0 \}$\,.
In Table~\ref{tab:tricriticalFPs} we list the anomalous dimension $\eta$, the $A$ function and the critical exponents $\theta_R$ values at the aforementioned fixed points. Some of these quantities can be found in Refs.~\cite{Jack:2015tka, Jack:2016utw, Osborn:2017ucf}.
\begin{figure}
\center
\includegraphics[width=.8\textwidth]{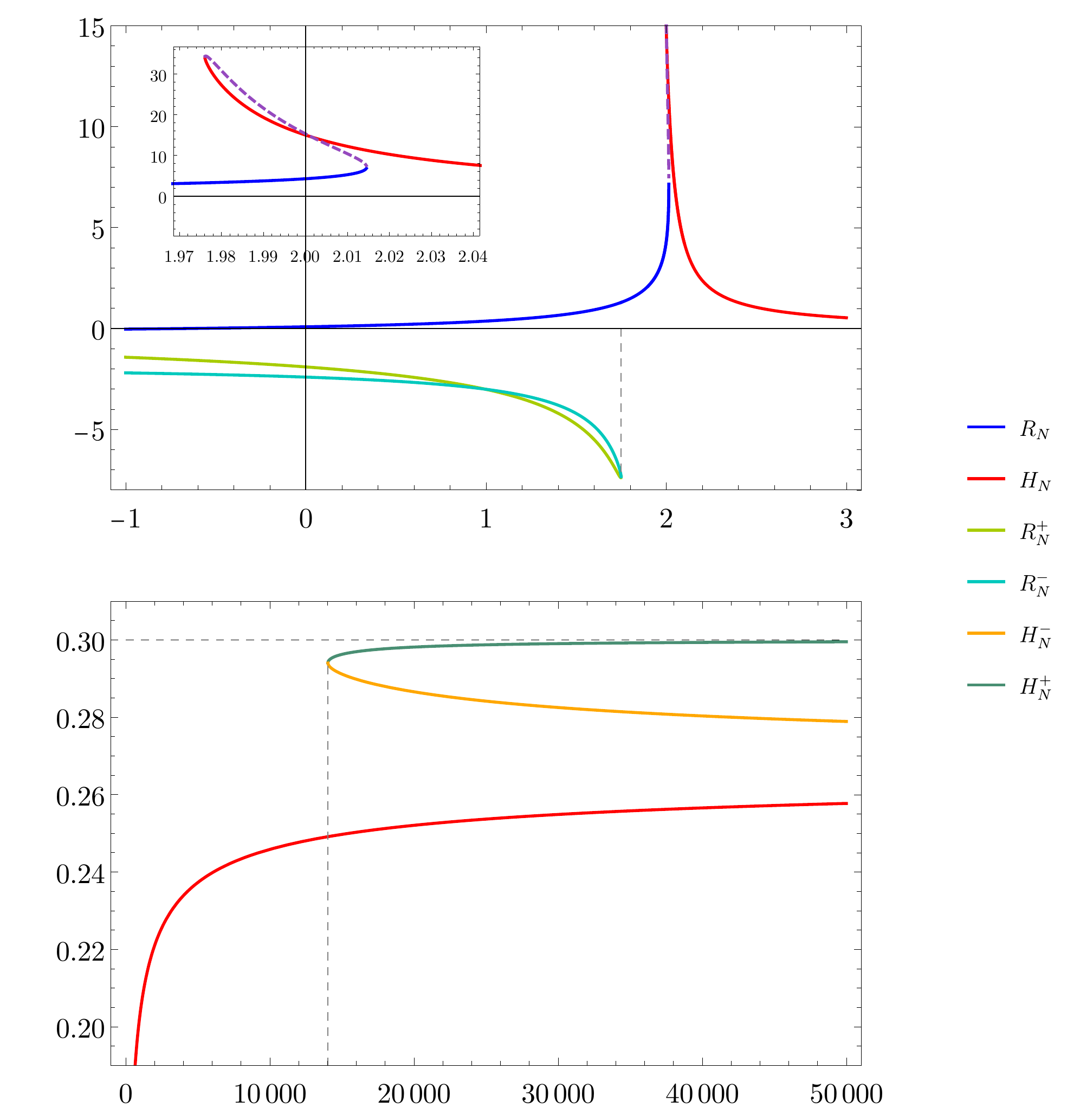}
\caption{In the top panel, we present the tricritical hypercubic fixed points (solutions of Eq.~\eqref{eq:Groebner}) in the range $-1 \leq N \leq 3$. In particular, we highlight the annihilation mechanism (inset) in terms of which we distinguish the solution $R_N$ (relevant for the discussion of the tricritical random universality classes in the $N\to 0$ limit) from the standard tricritical hypercubic one  $H_N$. In the bottom panel, instead, we present the tricritical hypercubic fixed points relevant to the discussion of the large $N$ limit behavior of the theory. Notice the pair $H_N^\pm$ appearing at $N = N^{(+)} \approx 14000$ that prevents the $H_N$ solution to approach the decoulped limit at $N\to \infty$.}
\label{Fig:tricriticalsFPs}
\end{figure}

We also find truly hypercubic fixed points which, due to their complexity, cannot be given in a closed analytical form.
However, one can elegantly express the hypercubic fixed point solutions as the roots of the following quintic polynomial
{\small
\begin{equation}\label{eq:Groebner}
\begin{split}
  \mathscr{P}(\sigma_3) & = \sigma_3 (10 \sigma_3-3) \left[4000\left(11664 N^4 - 114981 N^3+599698 N^2 - 1373948 N+1082776\right)\sigma_3^5 \right.\\
  & + 1200 \left(47208 N^4-1321709 N^3+10015426
   N^2-26585564 N+22877016\right) \sigma_3^4 \\
  & + 360  \left(2352 N^5 - 133790 N^4 - 3290479
   N^3 + 57207574 N^2 - 187161588 N + 175590440\right)\sigma_3^3 \\
  & - 108  \left(6552 N^5 + 134032 N^4 - 1432169 N^3 - 131870302 N^2 + 569826100 N - 587691368\right)\sigma_3^2\\
  & +810  \left(243
   N^5+11180 N^4+318148 N^3+5852320 N^2-26080672 N+28363776\right) \sigma_3\\
  & \left. - 6075 (3 N+2)
   \left(N^2+18 N-448\right)^2 \right] \,,
\end{split}
\end{equation}
}%
which is the first polynomial of the Gr\"obner basis of the set of equations $\beta_{\sigma_a}=0$. Solutions to $\mathscr{P}(\sigma_3)=0$
are in almost one-to-one correspondence to fixed points of the full system, because the couplings $\sigma_{1,2}$ are solved as functions of $\sigma_3$
through the other polynomials of the basis, that we do not give for brevity.
Notice that in Eq.~\eqref{eq:Groebner} the Gaussian and tricritical decoupled solutions are factorized
(the Gaussian solution $\sigma_3=0$ actually accounts for both the full Gaussian solution and for the tricritical $O(N)$
solution, which are distinguished from the other polynomials of the Gr\"obner basis). The remaining hypercubic solutions of \eqref{eq:Groebner} can be accessed numerically and the result is reported in Fig.~\ref{Fig:tricriticalsFPs}.
For $N<N^{(-)}\simeq \frac{7}{4}$ there are three possible fixed point solutions relevant to the discussion of the ``random'' tricritical universality class in the $N\to 0$ limit.
In agreement with the logic proposed in Ref.~\cite{Zinati:2020xcn} (for which one is led to consider only solutions such that $\sigma_3^\star>0$ in the limit), we identify $R_N$ as the only compatible `random' tricritical fixed point solution, at which we find $\eta = 0.00137842 \epsilon^2$.
While the other two solutions, namely $R_N^+$ and $R_N^-$, distinguished for values $N<N^{(-)}$, annihilate each other at $N^{(-)}$, the ``random'' solution $R_N$ ceases to exist at $N_R\simeq 2.015$. The mechanism behind this fact is reported in the inset of Fig.~\ref{Fig:tricriticalsFPs}, left panel. At $N_{H}\simeq 1.975$ a pair of fixed points emerge: one branch (dashed line in the inset) is responsible for annhilating the `random' solution $R_N$ at $N_R$, while the other branch $H_N^0$, which exists in all the range $N>N_H$ is the standard (tricritical) hypercubic one.
Concerning the large $N$ limit of the $\phi^6$-hypercubic theory, we notice that at $N^{(+)} \simeq 14000$, a pair of fixed points emerge, namely $H_N^+$ and $H_N^-$ and they become distinguished in the large $N$ limit \cite{Osborn:2017ucf}: while $H_N^-$ eventually annihilates the $H_N$ one, the fixed point $H_N^+$ tends to the tricritical Ising universality class which is characterized by $\sigma_3^\star=\frac{3}{10}$, see the discussion above.
This interesting mechanism shows that the large $N$ limit of the hypercubic solution $H_N$ does not correspond to the actual large $N$ limit behavior of the hypercubic theory at $d_c=3$, which is instead carried by the solution $H_N^+$.

Finally, for small values of $N$, specifically $N=1,2$, there are less independent monomials in the potential than for the general case $N\ge 3$, thus
separate discussions are required.
For $N=2$, the number of independent monomials in \eqref{eq:pot6} is reduced from three to two, because, for example, one can write the monomial
multiplying $\sigma_2$ as a linear combination of the other two. One can thus eliminate the $\sigma_2$ in favor of the other couplings, being left with the following system of reduced $\beta$-functions
\begin{align}
\beta_{\sigma_1}|_{N=2} & = \beta_{\sigma_1}|_{N=2, \sigma_2=0} + \frac{1}{3} \beta_{\sigma_2}|_{N=2, \sigma_2=0} \nonumber \\
& = -2 \sigma_1 \epsilon -
 \left(\frac{6976}{75}+\frac{2
 28 \pi ^2}{25}\right)
 \sigma_1^3+
 \left[\frac{112}{15} - \left(\frac{1072}{15}+\frac{33 \pi ^2}{5}\right)
 \sigma_3\right] \sigma_1^2 \nonumber \\
 & +
 \left[\frac{8\sigma_3}{3} - \left(\frac{284}{15}+\frac{3 \pi ^2}{2}\right)
 \sigma_3^2\right] \sigma_1 \,, \\
 \beta_{\sigma_3}|_{N=2} & = \beta_{\sigma_3}|_{N=2, \sigma_2=0} + \frac{2}{3} \beta_{\sigma_2}|_{N=2, \sigma_2=0} \nonumber \\
 & = -2 \sigma_3
 \epsilon -
 \left(\frac{12352}{75}+\frac{84 \pi ^2}{5}\right)\sigma_1^2\sigma_3 +
 \left[\frac{32
 \sigma_3}{3} - \left(\frac{3088}{15} + 21 \pi^2\right)
 \sigma_3^2\right]\sigma_1 \nonumber \\
 & -\left(\frac{1124}{15}+\frac{15 \pi^2}{2}\right)\sigma_3^3
 +\frac{20\sigma_3^2}{3} \,.
\end{align}
This system predicts only the tricritical Ising and the tricritical $O(2)$ universality classes
(we could have eliminated any other coupling, but the final RG system would have been the same, as we checked explicitly for this and all cases shown later).
This somehow reflects the critical case in $d_c=4$, for which the $N=2$ cubic theory predicts only the Ising and $O(2)$ fixed points (see Appendix \ref{sec:4-eps}).
We deduce that the nontrivial hypercubic fixed point solutions reported in Fig.~\ref{Fig:tricriticalsFPs} at $N=2$
are simply understood as analytical continuations of the two solutions $R_N$ and $H_N$, coming from $N=0$ and large-$N$, respectively.
Analogously, in the case $N=1$ the number of independent couplings is reduced to one, which we can choose to be $\sigma_1$
(thus eliminating $\sigma_2$ and $\sigma_3$). In this case the following reduced $\beta$-function
\begin{align}
\beta_{\sigma_1}|_{N=1} & = \beta_{\sigma_1}|_{N=1,\sigma_2=0, \sigma_3=0} \nonumber \\
& = - 2 \sigma_1
\epsilon - \left(\frac{1124}{15} +
\frac{15 \pi ^2}{2}\right)
 \sigma_1^3 + \frac{20\sigma_1^2}{3} \,,
\end{align}
predicts only the tricritical Ising universality class (see Table \ref{tab:tricriticalFPs} for the associated universal quantities).

Concerning the crossover exponents, from Table~\ref{tab:tricriticalFPs} we see that, as expected, at the tricritical isotropic $O(N)$ fixed point, the critical exponents associated with the deformations $X$ and $Y$ are identical and so they are the corresponding crossover exponents. At the tricritical Ising fixed point, instead, one can notice that $\theta_X = \theta_S$, while the quadratic deformation associated with $Y$ simply amounts to a shift in the critical temperature. We have also analysed the quadratic deformations at the hypercubic fixed points discussed above, which come from inserting the fixed point values in \eqref{eq:quadratic3d}. The final results are rather complicate and not particularly illuminating, so we do not report them here.

\section{Tetracritical model in \texorpdfstring{\bm{$d = \frac{8}{3}-\epsilon$}}{8/3-eps}}
\label{sec:8/3-eps}
The $\epsilon$-expansion below $d_c=\frac{8}{3}$ determines the properties of tetracritical fixed points. The effective potential in this case can be expressed following the line of reasoning in Section \ref{sec:hypercubicalgebra}, namely
\begin{equation}\label{eq:pot8}
\begin{split}
    V(\phi) & =  \tau_1\, \Bigl( \sum_i \phi_i^2 \Bigr)^4 +
    \tau_2\, \Bigl( \sum_i \phi_i^2 \Bigr)^2 \, \sum_j \phi_j^4 + \tau_3\, \sum_i \phi_i^2 \, \sum_j \phi_j^6 + \tau_4\, \Bigl( \sum_i \phi_i^4 \Bigr)^2 + \tau_5\, \sum_i \phi_i^8 \\
    & +  \sigma_1\, \Bigl( \sum_i \phi_i^2 \Bigr)^3  +	\sigma_2\, \sum_i \phi_i^2 \, \sum_j \phi_j^4 + \sigma_3\, \sum_i \phi_i^6 \\
    & + \lambda_1\, \Bigl( \sum_i \phi_i^2 \Bigr)^2 + \lambda_2\, \sum_i \phi_i^4 + m^2\, \phi_i^2 \,.
\end{split}
\end{equation}
The LO and NLO beta functionals $\beta_V$ and $\gamma_{ij}$ in this case are given by
\begin{equation}\label{eq:rg-tetra}
\begin{split}
        \beta_V  = &
        + \frac{1}{8} ~
        \begin{tikzpicture}[baseline=-.1cm]
        \draw (0,0) circle (.5cm);
        \draw (-.5,0) to [out=50,in=130] (.5,0);
        \draw (-.5,0) to[out=-50,in=-130] (.5,0);
        \filldraw [gray!50] (-.5,0) circle (2pt);
        \draw (-.5,0) circle (2pt);
        \filldraw [gray!50] (.5,0) circle (2pt);
        \draw (.5,0) circle (2pt);
        \end{tikzpicture}
        ~ +\frac{1}{160} ~
        \begin{tikzpicture}[baseline=-.1cm]
        \draw (1.5,0) circle (.5cm);
        \draw (1,0) to [out=50,in=130] (2,0);
        \draw (1,0) to [out=75,in=180] (1.5,.375);
        \draw (1.5,.375) to [out=0,in=115] (2,0);
        \draw (1,0) to [out=-75,in=180] (1.5,-.375);
        \draw (1.5,-.375) to [out=0,in=-115] (2,0);
        \draw (1,0) to [out=0,in=180] (2,0);
        \draw (1,0) to[out=-50,in=-130] (2,0);
        \filldraw [gray!50] (1.5,.5) circle (2pt);
        \draw (1.5,.5) circle (2pt);
        \filldraw [gray!50] (1,0) circle (2pt);
        \draw (1,0) circle (2pt);
        \filldraw [gray!50] (2,0) circle (2pt);
        \draw (2,0) circle (2pt);
        \end{tikzpicture}
        ~ + \frac{9}{80} ~
        \begin{tikzpicture}[baseline=-.1cm]
        \draw (3,0) circle (.5cm);
        \draw (2.531,-.171) to [out=-30,in=-150] (3.469,-.171);
        \draw (2.531,-.171) to [out=30,in=150] (3.469,-.171);
        \draw (3,.5) to[out=-45,in=115] (3.469,-.171);
        \draw (2.531,-.171) to[out=65,in=180] (3,.15);
        \draw (3,.15) to[out=0,in=115] (3.469,-.171);
        \draw (2.531,-.171) to [out=0,in=180] (3.469,-.171);
        \filldraw [gray!50] (3,.5) circle (2pt);
        \draw (3,.5) circle (2pt);
        \filldraw [gray!50] (3.469,-.171) circle (2pt);
        \draw (3.469,-.171) circle (2pt);
        \filldraw [gray!50] (2.531,-.171) circle (2pt);
        \draw (2.531,-.171) circle (2pt);
        \end{tikzpicture}
        ~ - \frac{3}{8} ~
        \begin{tikzpicture}[baseline=-.1cm]
        \draw (4.5,0) circle (.5cm);
        \draw (4.031,-.171) to [out=-30,in=-150] (4.969,-.171);
        \draw (4.031,-.171) to [out=30,in=150] (4.969,-.171);
        \draw (4.5,.5) to[out=-40,in=110] (4.969,-.171);
        \draw (4.969,-.171) to[out=150,in=-90] (4.5,.5);
        \draw (4.031,-.171) to [out=0,in=180] (4.969,-.171);
        \filldraw [gray!50] (4.5,.5) circle (2pt);
        \draw (4.5,.5) circle (2pt);
        \filldraw [gray!50] (4.969,-.171) circle (2pt);
        \draw (4.969,-.171) circle (2pt);
        \filldraw [gray!50] (4.031,-.171) circle (2pt);
        \draw (4.031,-.171) circle (2pt);
        \end{tikzpicture}\\
        & -\frac{\Gamma\left(\frac{1}{3}\right)^3}{24} ~
        \begin{tikzpicture}[baseline=-.1cm]
       \draw (6,0) circle (.5cm);
       \draw (5.531,-.171) to [out=-30,in=-150] (6.469,-.171);
       \draw (6,.5) to[out=-40,in=110] (6.469,-.171);
       \draw (6,.5) to[out=-90,in=30] (5.531,-.171);
       \draw (6,.5) to[out=220,in=70] (5.531,-.171);
       \draw (6.469,-.171) to[out=150,in=-90] (6,.5);
       \filldraw [gray!50] (6,.5) circle (2pt);
       \draw (6,.5) circle (2pt);
       \filldraw [gray!50] (6.469,-.171) circle (2pt);
       \draw (6.469,-.171) circle (2pt);
       \filldraw [gray!50] (5.531,-.171) circle (2pt);
       \draw (5.531,-.171) circle (2pt);
       \end{tikzpicture}
       ~ + \frac{3}{64} \left[\sqrt{3} \pi -3
          (2+\log 3)\right] ~
       \begin{tikzpicture}[baseline=-.1cm]
       \draw (7.5,0) circle (.5cm);
       \draw (7.031,-.171) to [out=-30,in=-150] (7.969,-.171);
       \draw (7.031,-.171) to [out=30,in=150] (7.969,-.171);
       \draw (7.5,.5) to[out=-90,in=30] (7.031,-.171);
       \draw (7.969,-.171) to[out=150,in=-90] (7.5,.5);
       \draw (7.031,-.171) to [out=0,in=180] (7.969,-.171);
       \filldraw [gray!50] (7.5,.5) circle (2pt);
       \draw (7.5,.5) circle (2pt);
       \filldraw [gray!50] (7.969,-.171) circle (2pt);
       \draw (7.969,-.171) circle (2pt);
       \filldraw [gray!50] (7.031,-.171) circle (2pt);
       \draw (7.031,-.171) circle (2pt);
       \end{tikzpicture} ~,\\
%
%     \beta_Z  = &
      \gamma_{ij}\phi_i\phi_j = &
       + ~ \frac{1}{2240} ~~
       \begin{tikzpicture}[baseline=-.1cm]
       \draw (1.5,0) circle (.5cm);
       \draw (1,0) to [out=50,in=130] (2,0);
       \draw (1,0) to [out=75,in=180] (1.5,.375);
       \draw (1.5,.375) to [out=0,in=115] (2,0);
       \draw (1,0) to [out=-75,in=180] (1.5,-.375);
       \draw (1.5,-.375) to [out=0,in=-115] (2,0);
       \draw (1,0) to [out=0,in=180] (2,0);
       \draw (1,0) to[out=-50,in=-130] (2,0);
       \filldraw [gray!50] (1,0) circle (2pt);
       \draw (1,0) circle (2pt);
       \filldraw [gray!50] (2,0) circle (2pt);
       \draw (2,0) circle (2pt);
       \end{tikzpicture} ~,
     \end{split}
\end{equation}
while a general expression for the $A$ function in $d=\frac{8}{3}-\epsilon$ is not yet available.
As the computation rapidly goes out of hand for increasing order, here we limit ourselves to report only the corresponding LO $\beta$-functions,
even though, in principle, the running is known to NLO in \eqref{eq:rg-tetra}. The $\beta$-function for the mass term reads,
\begin{align}
  \beta_{m^2} & = -2 m^2 + \frac{2}{15} \left[\left(N^2+6 N+8\right) \sigma_1 +3(N+4) \sigma_2 + 15 \sigma_3 \right] \lambda_1 \nonumber\\
  & + \frac{2}{15} \left[3 (N+4)\sigma_1 + (N+14)\sigma_2  + 15 \sigma_3\right] \lambda_2 \,,
\end{align}
while the ones for the quartic and sextic couplings are given by
\begin{align}
  \beta_{\lambda_1} &
  = - \frac{1}{3}(4+3\epsilon)\lambda_1 + \frac{1}{35} \left[2\left(N^2+10 N+24\right) \tau_1 + 5 N \tau_2 + 32 \tau_2 + 15 \tau_3+12 \tau_4\right] \lambda_1 \nonumber \\
  &
  +\frac{1}{35} \left[6 (N+8) \tau_1 + (N+28) \tau_2 + 15 \tau_3\right] \lambda_2 + \frac{2}{25} \left(N^2+18 N+56\right) \sigma_1^2  \nonumber\\
  &
  + \frac{2}{75} (N+50) \sigma_2^2 +\frac{12}{25}  (N+12) \sigma_1 \sigma_2 + \frac{12}{5} \sigma_1 \sigma_3 + \frac{4}{5} \sigma_2 \sigma_3 \,,
%\end{align}
%
\\
%
%\begin{align}
  \beta_{\lambda_2} &
  = - \frac{1}{3}(4+3\epsilon)\lambda_2 + \frac{1}{105} \left[\left(N^2\!+\!18 N\!+\!80\right)\tau_2  + 15 (N\!+\!10)\tau_3  + 6(N\!+\!28) \tau_4  + 210 \tau_5\right] \lambda_1 \nonumber \\
  &
  + \frac{1}{35} (16 \tau_1 + (N\!+\!40)\tau_2  + 55 \tau_3 + 2 \tau_4 N + 68 \tau_4 + 70 \tau_5) + \frac{4 (N\!+\!28)}{25} \sigma_2^2  \nonumber\\
  &
  +\frac{16}{25} (N\!+\!8) \sigma_1 \sigma_2 +\frac{48}{5} \sigma_1 \sigma_3 + \frac{56}{5} \sigma_2 \sigma_3 + 6 \sigma_3^2 \,,
\end{align}
and
\begin{align}
  \beta_{\sigma_1} & = -\frac{2}{3} (1 + 3\epsilon) \sigma_1 + \frac{3}{35} \left[2 \left(N^2+30 N+144\right) \tau_1 + 5 N \tau_2 + 92 \tau_2 + 15 \tau_3 + 12 \tau_4\right]\sigma_1 \nonumber\\
  &
  + \frac{1}{35} \left[18 (N+20) \tau_1 +(N+86) \tau_2 + 15 \tau_3\right]\sigma_2 + \frac{3}{7} (6 \tau_1+\tau_2)\sigma_3 \,,
%\end{align}
%
\\
%
%\begin{align}
  \beta_{\sigma_2} & = -\frac{2}{3} (1 + 3\epsilon) \sigma_2 + \frac{1}{35} (48 (N\!+\!13) \tau_1 + 3 (5 N \!+\! 204) \tau_2 + 360 \tau_3 + 2 N \tau_4 +232 \tau_4+70 \tau_5) \sigma_2 \nonumber\\
  & + \frac{1}{35} \left[\tau_2 \left(N^2\!+\!78 N\!+\!680\right)+15 \tau_3 (N\!+\!40)+6 \tau_4 (N\!+\!88)+210 \tau_5\right] \sigma_1 \nonumber\\
  &
  + \frac{3}{7} (48 \tau_1+29 \tau_2+15 \tau_3+2 \tau_4) \sigma_3 \,,
%\end{align}
%
\\
%
%\begin{align}
  \beta_{\sigma_3} & = -\frac{2}{3} (1 + 3\epsilon) \sigma_3 + \frac{1}{35} \left[16(N\!+\!20) \tau_2  + 15 (N\!+\!44)\tau_3 + 816 \tau_4 + 980 \tau_5\right]\sigma_2 \nonumber\\
  & + \frac{6}{7} \left[(N\!+\!12)\tau_3 + 16 \tau_4 + 28 \tau_5\right]\sigma_1 + \frac{3}{7} (16 \tau_1+40 \tau_2+55 \tau_3+68 \tau_4+70 \tau_5)\sigma_3 \,.
\end{align}

Finally the $\beta$-functions for the marginal couplings read
\begin{align}
  \beta_{\tau_1} & = - 3 \tau_1
  \epsilon + \frac{1}{70} \left[4 \left(3 N^2+150 N+1072\right)\tau_1^2 + 12(5N + 152 ) \tau_1 \tau_2 + 180 \tau_1 \tau_3 + 144 \tau_1\tau_4  \right.\nonumber\\
   & \left. + (N  + 144) \tau_2^2 + 30 \tau_2 \tau_3\right] \,, \\
%   & \nonumber\\
   \beta_{\tau_2} & = -3
   \tau_2 \epsilon + \frac{1}{70} \left[4 \left(N^2+138 N+1840\right) \tau_1
   \tau_2 + 60(N + 70) \tau_1 \tau_3 + 24 (N+148) \tau_1 \tau_4 \right. \nonumber\\
   & \left. + 840 \tau_1 \tau_5 + 2 (17 N\!+\!968)\tau_2^2 + 1360 \tau_2 \tau_3 + 4 (N\!+\!204) \tau_2\tau_4 + 140 \tau_2 \tau_5 + 225 \tau_3^2 + 60 \tau_3 \tau_4 \right] \,, \\
%   & \nonumber\\
   \beta_{\tau_3} & = -3 \tau_3 \epsilon + \frac{1}{105} \left[120 (3 N+64) \tau_1 \tau_3 + 5760 \tau_1 \tau_4 + 10080 \tau_1 \tau_5 + 128 (N+25) \tau_2^2 \right. \nonumber\\
   & \left. + 30 (3 N+268) \tau_2 \tau_3 + 5856 \tau_2 \tau_4 + 5880 \tau_2\tau_5 + 3075 \tau_3^2 + 3600 \tau_3 \tau_4 + 3150 \tau_3 \tau_5 + 384
   \tau_4^2\right] \,,\\
%   & \nonumber\\
   \beta_{\tau_4} & = - 3 \tau_4 \epsilon + \frac{1}{420} \left[13440 \tau_1 \tau_4 + 2 \left(N^2+98 N+1600\right) \tau_2^2 + 60 (N+50) \tau_2 \tau_3 \right. \nonumber \\
   & \left. + 24 (N+468) \tau_2 \tau_4 + 840 \tau_2 \tau_5 + 225 \tau_3^2 + 5160
   \tau_3 \tau_4 + 24 (N+176) \tau_4^2 + 1680 \tau_4\tau_5\right] \,,\\
%   & \nonumber\\
   \beta_{\tau_5} & = -3 \tau_5 \epsilon + \frac{1}{140} \left[4480 \tau_1 \tau_5 + 160 (N+30) \tau_2\tau_3 + 6400 \tau_2 \tau_4 + 11200 \tau_2 \tau_5 + 75 (N+68)\tau_3^2\right. \nonumber \\
   & \left. + 12960 \tau_3 \tau_4 + 15400 \tau_3 \tau_5 + 7872
   \tau_4^2 + 19040 \tau_4 \tau_5 + 9800 \tau_5^2\right] \,.
\end{align}
The anomalous dimension is obtained from diagonalizing $\gamma_{ij}$ with an additional factor two, as usual, and, in this case, it reads
\begin{align}
  \eta &
  = \frac{2}{3675} \left(N^3+12 N^2+44 N+48\right) \tau_1^2 + \frac{1}{3675}(N^2+33 N+176) \tau_2^2 + \frac{1}{490} (N+27) \tau_3^2 \nonumber\\
  &
  + \frac{2}{1225} (N+34) \tau_4^2 + \frac{2}{35} \tau_5^2 + \frac{4}{1225} \left(N^2+10 N+24\right) \tau_1 \tau_2 + \frac{4}{245} (N+6) \tau_1 \tau_3
  \nonumber\\
  &
  + \frac{4}{1225}  (3 N+32) \tau_1 \tau_4 + \frac{2}{245} (N+13) \tau_2 \tau_3  + \frac{4}{1225}  (N+34) \tau_2 \tau_4 \nonumber\\
  & + \frac{4}{35} \left[(\tau_1 + \tau_2  + \tau_3 + \tau_4)\tau_5 + \tau_3 \tau_4 \right] \,.
\end{align}

The system of $\beta$-functions admits a total of $32$ fixed points, including real and complex ones.
However, as in the tricritical case, we restrict ourselves to consider critical theories which are real and perturbatively unitary,
thus we limit our attention only to real fixed points with positive anomalous dimension $\eta>0$ and real spectrum
(the spectrum is real for all the operators of the basis of the operators that we included to the potential, we cannot exclude that irrelevant
operators come with imaginary parts).
We then require the fixed point potential \eqref{eq:pot8} to be stable, namely, bounded from below: this amounts at considering fixed points for which
the sum of the marginal couplings is positive,
$(\tau^{\star}_1 + \tau^{\star}_2 +  \tau^{\star}_3 + \tau^{\star}_4 + \tau^{\star}_5 ) > 0$.
These requirements boil down the plethora of fixed points to a manageable subset. Apart the trivial Gaussian fixed point, the system of $\beta$-functions admits the tetracritical Ising fixed point with coordinates given by
$\{\tau^{\star}_1 = 0 , \tau^{\star}_2 = 0 , \tau^{\star}_3 = 0, \tau^{\star}_4 = 0, \tau^{\star}_5 = \frac{3}{70} \epsilon\}$ and LO anomalous dimension $\eta = \frac{9 \epsilon ^2}{85750}$ \cite{Zinati:2019gct}.
We then find a fixed point representing the tetracritical $O(N)$ universality class with coordinates
$\{\tau^{\star}_1 = \frac{105 \epsilon }{2 \left(3 N^2+150 N+1072\right)} , \tau^{\star}_2 = 0 , \tau^{\star}_3 = 0, \tau^{\star}_4 = 0, \tau^{\star}_5 = 0 \}$\, and LO anomalous dimension given
by $\eta = \frac{3 (N+2) (N+4) (N+6) \epsilon ^2}{2 (3 N^2+150 N+1072)^2}$. The critical exponents of the tetracritical $O(N)$ universality class, for the specific values $N=2,3,4$, can be checked against Ref.~\cite{Zinati:2019gct}, while the limit $N\to0$, relevant to the discussion of the tetracritical self-avoiding-walks (SAW) universality class has been given in Ref.~\cite{Zinati:2020xcn}.

\begin{figure}
\center
\includegraphics[width=.8\textwidth]{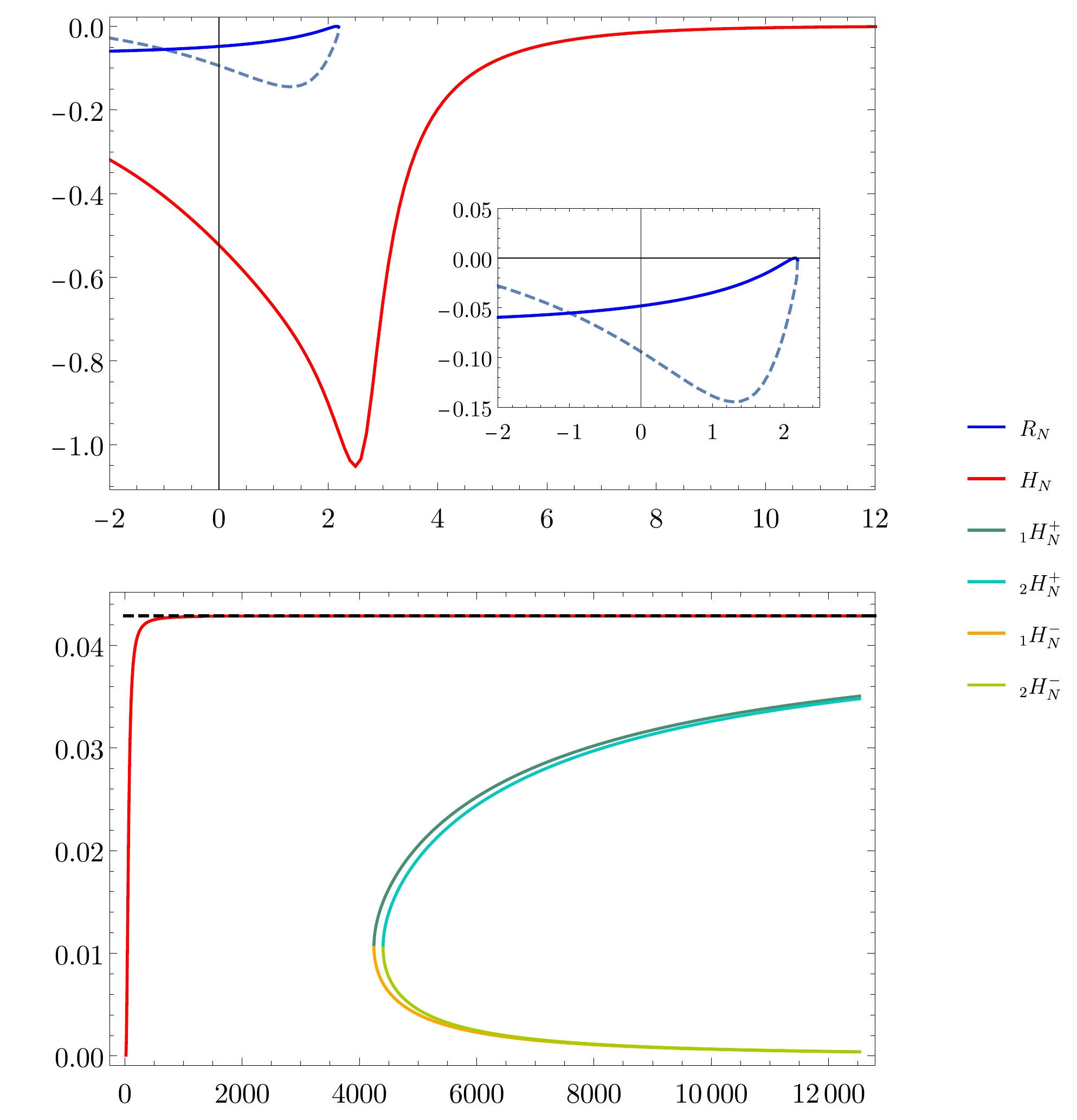}
\caption{In the top panel, we present the fixed point value $\tau_5^\star(N)$  at the tetracritical hypercubic solutions, in the range $-2 \leq N \leq 12$. In particular, we highlight the annihilation mechanism (inset) in terms of which we distinguish the solution $R_N$ ceases to exist. In the bottom panel, instead, we present the tetracritical hypercubic fixed points relevant to the discussion of the large $N$ limit behavior of the theory.}
\label{Fig:tetracriticalsFPs}
\end{figure}

Additionally, we still find several non-trivial fixed points with true hypercubic symmetry, that,
because of their complexity, we are not able to express neither in a closed analytical form, nor as roots of a polynomial of the Gr\"obner basis
like in the previous section.
However, they can still be accessed numerically, which will be our main tool of investigation from this point forward.
We find it useful to report the resulting scenario in terms of the fixed-point coupling $\tau^{\star}_5$ as a function of the number of components $N$,
similarly to what we have done in the previous section. The solutions are fixed points of the full system of five beta functions continued in $N$,
but, as previously discussed, it is necessary to consider the special cases $N=1,2,3$ separately, because there are less independent monomials
in the respective bases. The result is thus an analytic continuation in $N$ and is shown in Fig.~\ref{Fig:tetracriticalsFPs}.

We immediately recognize a $\phi^8$-hypercubic solution with $H_N$ symmetry that exists in the whole the range of $N$
with the desired physical properties for $N > 3$, that is denoted $H_N$ likewise the group in Fig.~\ref{Fig:tetracriticalsFPs}.
In the large $N$ limit, this solution tends asymptotically to the decoupled tetracritical Ising universality class,
for which we recall $\tau^{\star}_5=\frac{3\epsilon}{70}$,
so the hypercubic spins decouple in the limit (which is a rather different behavior in comparison to the tricritical model that did not decouple).
At the values $N_1 = 4260$ and $N_2 = 4410$, two new pairs of fixed points emerge, respectively $_1H^\pm_N$ and $_2H^\pm_N$.
In the large $N$ limit, the upper branches $_1H_N^+$ and $_2H_N^+$ tend to $H_N$ as far as our numerics can tell
(asymptotically to the tetracritical decoupled Ising universality class), instead the lower branches $_1H_N^-$ and $_2H_N^-$ tend to zero.
The pairs do not affect the large-$N$ limit of the $H_N$ solution because they emerge ``below'' it, using the coupling $\tau_5$ as reference (see the right panel of Fig.~\ref{Fig:tetracriticalsFPs}).

It is also interesting to discuss the fate of the solution $H_N$ in the limit $N\to 0$, in relation to the randomly diluted tetracritical model.
Similarly to the tricritical case, we notice that the multicritical solution $H_N$ does not correspond to the $N\to 0$ fixed point that
has been argued to describe the ``random'' tetracritical universality class identified in Ref.~\cite{Zinati:2020xcn}
on the basis of physical requirements that hold in relation to the replica approach to describe the diluted system.
We denote the continuation in $N$ of the random fixed point as $R_N$ to make the distinction with $H_N$ clear.
In the left panel of Fig.~\ref{Fig:tetracriticalsFPs} one can see that $R_N$ ceases to exist at $N_R \simeq 2.2$ by annihilating with a not-physical fixed point displayed with a dashed line.

Finally we must discuss separately the cases $N=1,2,3$ separately, because in these three cases
the five hypercubic terms in \eqref{eq:pot8} are not independent.
For $N=1$, the system can be reduced to a single invariant, namely $V(\phi) = \tau_1 \phi^8$,
which trivially describes only the tetracritical Ising universality class (with fixed point $\tau_1^{\star} = \frac{12\epsilon}{35}$).
For $N=2$, we find only three independent couplings: one can eliminate $\tau_2$ and $\tau_3$
in favor of the other three couplings being left with the following system of reduced $\beta$-functions
{\small
\begin{align}
\beta_{\tau_1}|_{N=2} & = \beta_{\tau_1}|_{N=2, \tau_2=0, \tau_3=0} + \frac{1}{2}\beta_{\tau_2}|_{N=2, \tau_2=0, \tau_3=0} + \frac{1}{4} \beta_{\tau_3}|_{N=2, \tau_2=0, \tau_3=0} \nonumber \\
& = -3 \tau_1 \epsilon + \frac{346
 \tau_1^2}{35}+\tau_1 \left(\frac{363
 \tau_4}{70}+\frac{15
 \tau_5}{4}\right)
 +\frac{4 \tau_4^2}{35} \,, \\
 \beta_{\tau_4}|_{N=2} & = \beta_{\tau_4}|_{N=2, \tau_2=0, \tau_3=0} - \frac{1}{2}\beta_{\tau_2}|_{N=2, \tau_2=0, \tau_3=0} - \frac{3}{4} \beta_{\tau_3}|_{N=2, \tau_2=0, \tau_3=0} \nonumber \\
 & = -3 \tau_4 \epsilon + \tau_1 \left(-\frac{61
 \tau_4}{14}-\frac{39
 \tau_5}{4}\right)+\frac{13
 \tau_4^2}{14}+\frac{\tau_4
 \tau_5}{2} \,, \\
\beta_{\tau_5}|_{N=2} & = \beta_{\tau_5}|_{N=2, \tau_2=0, \tau_3=0} + \beta_{\tau_2}|_{N=2, \tau_2=0, \tau_3=0} + \frac{3}{2} \beta_{\tau_3}|_{N=2, \tau_2=0, \tau_3=0} \nonumber \\
& = -3
\tau_5 \epsilon +  \tau_1
 \left(\frac{117 \tau_4}{7}+\frac{47
 \tau_5}{2}\right)+\frac{54\tau_4^2}{7}+17 \tau_4
 \tau_5+\frac{35 \tau_5^2}{4}\,.
\end{align}
}%
This system, similarly to the critical and tricritical cases for $N=2$, predicts only the tetracritical decoupled Ising fixed point and the tetracritical-$O(2)$ universality classes, the latter being characterised by $\eta=\frac{9\epsilon^2}{59858}$
(in other words, there does not seem to any multicritical theory of spins on a square, according to our analysis).
The last special case in $N=3$, in which there are four independent couplings. Eliminating $\tau_3$ in favor of the others, we are led to the following system of reduced $\beta$-functions
{\small
\begin{align}
 \beta_{\tau_1}|_{N=3} & = \beta_{\tau_1}|_{N=3, \tau_3=0} - \frac{1}{8} \beta_{\tau_3}|_{N=3, \tau_3=0} \nonumber \\
 & = -3 \tau_1 \epsilon + \frac{1549 \tau_1^2}{140}+\tau_1 \left(\frac{501
   \tau_2}{140}-\frac{3 \tau_4}{5}-\frac{3
   \tau_5}{2}\right) -\frac{13
   \tau_2^2}{48}+\tau_2 \left(-\frac{61
   \tau_4}{70}-\frac{7 \tau_5}{8}\right)-\frac{2
   \tau_4^2}{35} \,, \\
 \beta_{\tau_2}|_{N=3} & = \beta_{\tau_2}|_{N=3, \tau_3=0} + \frac{3}{4} \beta_{\tau_3}|_{N=3, \tau_3=0} \nonumber \\
 & = -3
 \tau_2 \epsilon + \tau_1 \left(\frac{2263 \tau_2}{140}+\frac{813
   \tau_4}{70}+\frac{21 \tau_5}{2}\right)+\frac{383
   \tau_2^2}{56}+\tau_2 \left(\frac{939
   \tau_4}{140}+\frac{11 \tau_5}{2}\right) +\frac{12 \tau_4^2}{35} \,, \\
 \beta_{\tau_4}|_{N=3} & = \beta_{\tau_4}|_{N=3, \tau_3=0} - \frac{3}{8} \beta_{\tau_3}|_{N=3, \tau_3=0} \nonumber \\
 & = -3
 \tau_4 \epsilon + \tau_1 \left(\frac{10 \tau_4}{7}-\frac{9
   \tau_5}{2}\right)-\frac{157
   \tau_2^2}{336}+\tau_2 \left(\frac{3
   \tau_4}{4}-\frac{19 \tau_5}{8}\right)+\frac{31
   \tau_4^2}{28}+\frac{\tau_4 \tau_5}{2} \,,\\
 \beta_{\tau_5}|_{N=3} & = \beta_{\tau_5}|_{N=3, \tau_3=0} + \frac{3}{4} \beta_{\tau_3}|_{N=3, \tau_3=0} \nonumber \\
 & = -3
 \tau_5 \epsilon + \tau_1 \left(\frac{36 \tau_4}{7}+13
   \tau_5\right)+\frac{16 \tau_2^2}{5}+\tau_2
   \left(\frac{383 \tau_4}{35}+\frac{61
   \tau_5}{4}\right)+\frac{258 \tau_4^2}{35}+17
   \tau_4 \tau_5+\frac{35 \tau_5^2}{4} \,.
\end{align}
}%
This system predicts the tetracritical Ising fixed point, the tetracritical $O(3)$ fixed point characterized by $\eta = \frac{945\epsilon^2}{4798802}$
and a genuinely $\phi^8$-hypercubic solution $H_{N=3}$ with anomalous dimension given by $\eta = 0.000139712 \, \epsilon^2$.

The fixed point of the general $\phi^8$-hypercubic solution $H_N$, which is plotted in Fig.~\ref{Fig:tetracriticalsFPs}, is therefore not representative of a truly hypercubic behavior for the special cases $N=1,2$ as the analysis of the reduced systems of $\beta$-functions reveals,
but is purely understood as an analytical continuation. If one chooses to study the continuation of the solution to the special cases $N=1,2$,
it is easy to see that, because of the redundancy of the basis of operators, they are fixed points with pseudo-Gaussian features and their spectrum
coincides with either the $O(N)$ or the decoupled tetracritical case. As far as our analysis can tell,
we do not find a genuinely hypercubic fixed point for the interesting value $N=2$,
and the first natural ones appear at $N=3$.

\section{Conclusion}
\label{sec:conclusions}

This paper could be considered part of a series of works \cite{Zinati:2017, Zinati:2019gct, Safari:2020eut, Codello:2020mnt}
that share the common theme of discussing and classifying ``discrete'' universality classes, namely renormalization group critical points of scalar field theories which are characterised by a finite discrete symmetry group.
In particular, here we concentrated on the multicritical behavior of scalar field theories endowed with the symmetry of the hypercubic point group $H_N$. From a group-theoretical point of view, hypercubes (as well as hypertetrahedra) belong to a family of regular polytopes that exist in $\mathbb{R}^N$,
with $N$ indicating the number of independent components $\phi_i$ of the associated order parameter. It is also possible to analytically continue $N$, which makes the results particularly interesting in connection to the study of its critical behavior as a function of $N$, which makes further universal properties accessible. For example, a notable limit is $N\to 0$ of the hypercubic theory, which is known to be representative of the universality class of randomly diluted spin systems, for which some of us recently investigated multicritical effects of arbitrary distributions of the disorder \cite{Zinati:2020xcn}. In this paper we moved forward to a systematic analysis of the multicritical behavior of hypercubic scalar field theories as a function of $N$, paying particular attention to some non-trivial annihilation mechanisms involved in the $N\to 0$ and $N\to \infty$ limits.

To assess the continuation in $N$ correctly, we first reviewed the algebra of hypercubic invariant polynomials to determine the most general basis for the interaction potentials. This must be done with care, because, for a given multicritical interaction and finite and small $N$, the counting of operators is not general,
because a limited number of parameters $\phi_i$ reduces the total number of independent interactions.
This analysis is not restricted to a unique quadratic polynomial, but instead we considered the quadratic deformations associated with irreducible representations of the hypercubic group. This is relevant to the discussion of the crossover exponents that characterize the critical behavior of systems which depend on the interplay of more than one order parameter. Besides the general quadratic deformations, we also included all possible relevant $H_N$-singlets in the potentials,
which give further informations on the universal properties and can be used to determine some of the conformal data of the underlying CFT \cite{Codello:2017a}.

On general grounds, the discussion of the multicritical behavior of scalar field theories with $\phi^{2n} $ interactions involves, perturbatively, the renormalization group analyses and an $\epsilon$-expansion from rational space-time dimensions. Within the functional perturbative RG, this analysis can be carried out systematically \cite{ODwyer:2007, Codello:2017a}. Most of the RG equations that we adopted can be derived from simpler single component equations, so we provide the theoretical rationale behind the fact that the single-component beta functionals can be directly enhanced to their multicomponent counterpart.

We have focused on two upper critical dimensions, namely $d_c=3$ and $d_c=8/3$, corresponding respectively to a $\phi^6$ and a $\phi^8$ scalar field theory,
though our analysis could be generalized to higher rank interactions at the cost of numerical complexity. In both cases, our analysis revealed a rich spectrum of fixed points which, in turn, determine the non-trivial mechanism occurring behind the large $N$ as well the $N\to 0$ limit of the theories.
In particular, in $d=3-\epsilon$ our analysis showed that the ``tricritical'' $\phi^6$-hypercubic fixed point solution, denoted $H_N$ as the group, that interpolates the limits $N=2$ and $N=3$,  does not coincide with the continuation of the ``random'' universality class, denoted $R_N$,  that interpolates with the $N\to 0$ limit given in Ref.~\cite{Zinati:2020xcn}. A marked difference is that the ``random'' model fixed point solution exists only for $N\lesssim2.015$.
On the opposite side, the analysis of the large $N$ limit reveals that the standard hypercubic solution $H_N$ annihilates asymptotically with the lower branch $H_N^-$ of a pair of fixed point $H_N^\pm$ that emerge from a complex conjugate pair at $N\simeq 14000$. The upper branch  $H_N^+$ instead tends asymptotically to the tricritical decoupled Ising universality class. In other words, the large-$N$ limit of the tricritical hypercubic model does not
tend to ($N$ copies of) the decoupled tricritical Ising model, which straightforwardly implies that if one were to construct a large-$N$ expansion in fixed dimension by seeding the decoupled model solution, such expansion would fail. Notice that this does not happen with the standard critical hypercubic model,
that in the large-$N$ limit tends to the decoupled Ising.

In the case of $d=8/3-\epsilon$, we identified a ``tetracritical'' $\phi^8$-hypercubic solution, also denoted $H_N$, that in all the whole range $N \geq 0$.
Differently from the tricritical solution, in the large $N$ limit the tetracritical solution tends to the tetracritical decoupled Ising universality class
In the $N\to 0$ limit, instead, it is still distinguished from the ``random'' tetracritical universality, again denoted $R_N$, that for $N=0$ was give in Ref.~\cite{Zinati:2020xcn}. The tetracritical random solution exists for $N\lesssim2.2$.
The large-$N$ limit of the tetracritical model can reach the decoupled solution asymptotically because there are no pairs of complex conjugate solutions
interrupting its path.
In fact, we see numerically two pairs of fixed point solutions, $_1H^\pm_N$ and $_2H^\pm_N$, emerging from complex conjugate pairs at some large values of $N$, but they do not intercept the $H_N$ solution. They still participate non-trivially to the large $N$ limit: the upper branches $_1H_N^+$ and $_2H_N^+$ of these solutions tend to the same large $N$ limit of $H_N$, that is, to the tetracritical Ising universality class, while the two lower branches tend to zero,
as far as our numerical analysis can tell.

One speculation that we could make, based on the first few multicritical hypercubic models that we considered in this work,
is that not all $\phi^{2n}$-hypercubic model have the corresponding decoupled model as large-$N$ limit. It could be that
only for $n$ \emph{even}, the hypercubic theories decouple in the limit, while for $n$ \emph{odd} they do not.
The mechanism preventing the decoupling of the spins in the limit would be the birth, at large values of $N$, of pairs of fixed points,
that collide asymptotically with the hypercubic solution. This mechanism is obviously confirmed by our findings for $n=2,3,4$.
One general strategy to try to prove it could follow our procedure: use the Gr\"obner basis to find a polynomial in the coupling
that weights the (sum of the) decoupled interaction and test the behavior of the roots of this polynomial.
The application of this idea is essentially what gives us Figs.~\ref{Fig:tricriticalsFPs}~and~\ref{Fig:tetracriticalsFPs} for the special cases $n=3,4$.
Similar nontrivial behaviors in the large-$N$ limits of $O(N)$ multicritical models has been already observed,
and it also depends on the value of $n$ being either even or odd \cite{henriksson2020analytic, Gracey:2017okb, Eyal:1996da, Defenu:2020cji}.

We discussed separately the RG equations of the $\phi^6$- and $\phi^8$-hypercubic models for the special values $N=1,2$
and $N=1,2,3$, respectively.
The reason is that for these values
the number of independent operators is less than the case of arbitrary $N$ (for example, for $N=1$ there is only one scalar, so only one independent marginal interaction can be formed, but similar simplifications occur for $N=2,3$).
The presence of a degeneracy in the number of independent invariant monomials, as seen from the point of view of the RG equations for arbitrary $N$, makes it such that the general $N$ fixed points see pseudo-Gaussian solutions for the lower $N$ cases (the spectrum has a mixture of Gaussian features
and nontrivial ones that cancel out if the correct bases are taken into account).
To truly explore the cases $N=1,2,3$, we have found the correct combinations of the general RG equations that gives the reduced (independent) system
of $\beta$-functions. Our analysis reveals that only in $d_c=\frac{8}{3}$ and for $N=3$ the reduced system
of $\beta$-functions predicts a true hypercubic fixed-point. All the other cases, instead, revealed no fixed-point solution associated with a truly hypercubic symmetry, but only solutions corresponding to multicritical Ising and $O(N)$ universality classes. In particular, in the models that we considered
there is no fixed point with the symmetries of a square and $N=2$.

For completeness, we reviewed the case of $d_c=4$ (in appendix~\ref{sec:4-eps}) for which, as a new result, we give the NNLO contributions to the $A$-function in appendix~\ref{sec:4-eps}. We have take the opportunity to discuss the behavior of the quadratic deformations at the hypercubic fixed point, commenting explicitly the physical mechanism occurring when two deformations coincide (which is known to be related to logarithmic CFTs \cite{Cardy:1999, Cardy:2013, Zinati:2020xcn}).
Finally, the inclusion of a singlet in the fields' multiplet allows constructing the interactions with an odd number of fields. In this case, we limited ourselves to report the beta functions of the $\phi^5$-hypercubic theory in $d_c=10/3$ given in appendix~\ref{sec:10/3-eps}, which are relevant for the discussion of a generalization of the random universality class presented in Ref.~\cite{Zinati:2020xcn}.

A future prospect of our approach could be to attempt the construction of multicritical $MN$-model,
that are theories with $ O(m)^n\rtimes S_n$ symmetry, which can be seen as a natural generalization of hypercubic symmetry \cite{mukamel1975, SHPOT1, SHPOT2}.
It is especially interesting to attempt this generalization because $MN$-models are seeing a resurging interest,
mostly coming from the conformal bootstrap literature, which is motivated by the possible presence of two distinct universality classes
that could have direct experimental implications \cite{Henriksson1, Henriksson2}.

\section*{Acknowledgements}
RBAZ is grateful to Matthieu Tissier for useful comments. OZ thanks Johan Henriksson for useful discussions.

\paragraph{Funding information}
RBAZ acknowledges the support from the French ANR through the project NeqFluids (grant ANR-18-CE92-0019).

\clearpage

\appendix

\section{Critical models in \texorpdfstring{\bm{$d = 4-\epsilon$}}{4-eps}}
\label{sec:4-eps}

In this appendix we review the known case of the hypercubic model in $d=4-\epsilon$.
The critical properties of the system below $d_c = 4$ can be described in terms of the following potential \cite{Aharony:1973a, Aharony:1973b, Wallace:1973},
\begin{equation}
\label{eq:pot4}
  V(\phi) =  m^2\, \phi_i^2 + \lambda_1\, \Bigl( \sum_i \phi_i^2 \Bigr)^2 + \lambda_2\, \sum_i \phi_i^4 \,,
\end{equation}
which can be obtained following the logic discussed in Section \ref{sec:hypercubicalgebra}.
The renormalized model predicts four fixed points: the Gaussian one, the Ising one in which the $N$ components of the field decouple
(therefore referred to as decoupled Ising), the $O(N)$-symmetric and the hypercubic fixed point. The three non-trivial  fixed points have been extensively studied in the RG literature: see for example Refs.~\cite{Batkovich:2016, Kompaniets:2017, Adzhemyan:2019gvv, Carmona:1999} for present-day six loops order results within the $\epsilon$-expansion.

There has been special interest in the determination of the stability properties of the aforementioned fixed points: it was shown that while the Gaussian and Ising fixed points are always unstable for arbitrary values of $N$, the $O(N)$-symmetric, referred to as isotropic, and the cubic, referred to anisotropic, compete with each other and their respective stability depends on $N$. For $N<N_c$, where $N_c$ is some critical value that depends on $d$, the isotropic fixed point is stable while for $N>N_c$ the opposite is true. The long debated issue is the value of $N_c$ for a three-dimensional system, $d=3$. To our best knowledge, the reference field-theoretical estimate is still provided by six-loop results of Ref.~\cite{Carmona:1999} and recently Ref.~\cite{Adzhemyan:2019gvv} which predict $N_c\simeq 2.9$ and strengthen the existing arguments in favor of the stability of the cubic fixed point in the physical case $N=3$.
We mention here that, interestingly, the authors of Refs.~\cite{Stergiou:2018gjj, Kousvos:2018rhl, Kousvos:2019hgc} claim evidence of a new $3d$ CFT characterized by the cubic symmetry group which is different from the one obtainable by standard $\epsilon$-expansion starting from $d=4-\epsilon$. This fact certainly deserves attention and could possibly be investigated by means of the functional non-perturbative RG extending the analysis carried out in Ref.~\cite{Tissier:2002zz}.
The determination of $N_c$, as well as the assessment of the existence of this new cubic $3d$ CFT, are, however, not the scope of the present paper.

In the functional perturbative RG, the LO, NLO and NNLO contributions to the potential's $\beta$-functional can be expressed diagrammatically as
\begin{equation}\label{BBetasd=4}
\begin{split}
        \beta_V  = &
        + \frac{1}{2} ~
        \begin{tikzpicture}[baseline=-.1cm]
        \draw (0,0) circle (.5cm);
        \filldraw [gray!50] (.5,0) circle (2pt);
        \draw (.5,0) circle (2pt);
        \filldraw [gray!50] (-.5,0) circle (2pt);
        \draw (-.5,0) circle (2pt);
        \end{tikzpicture}
        ~ -\frac{1}{2} ~
        \begin{tikzpicture}[baseline=-.1cm]
        \draw (1.5,0) circle (.5cm);
        \draw (1,0) to[out=0,in=180] (2,0);
        \filldraw [gray!50] (1.5,.5) circle (2pt);
        \draw(1.5,.5) circle (2pt);
        \filldraw [gray!50] (1,0) circle (2pt);
        \draw (1,0) circle (2pt);
        \filldraw [gray!50] (2,0) circle (2pt);
        \draw (2,0) circle (2pt);
        \end{tikzpicture}
        ~ -\hspace{2pt}\frac{1}{8} \hspace{2pt} ~
        \begin{tikzpicture}[baseline=-.1cm]
        \draw (3,0) circle (.5cm);
        \draw (3,.5) to[out=-90,in=180] (3.5,0);
        \filldraw [gray!50] (3,.5) circle (2pt);
        \draw(3,.5) circle (2pt);
        \filldraw [gray!50] (3.5,0) circle (2pt);
        \draw (3.5,0) circle (2pt);
        \draw (2.5,0) to[out=0,in=90] (3,-.5);
        \filldraw [gray!50] (2.5,0) circle (2pt);
        \draw (2.5,0) circle (2pt);
        \filldraw [gray!50] (3,-.5) circle (2pt);
        \draw(3,-.5) circle (2pt);
        \end{tikzpicture}
        ~ +\frac{\zeta_3}{2} ~
        \begin{tikzpicture}[baseline=-.1cm]
        \draw (4.5,0) circle (.5cm);
        \draw (4.5,0) to [out=90,in=-90] (4.5,.5);
        \draw (4.5,0) to[out=-30,in=150] (4.969,-.171);
        \draw (4.5,0) to[out=210,in=30] (4.031,-.171);
        \filldraw [gray!50] (4.5,.5) circle (2pt);
        \draw (4.5,.5) circle (2pt);
        \filldraw [gray!50] (4.5,0) circle (2pt);
        \draw (4.5,0) circle (2pt);
        \filldraw [gray!50] (4.969,-.171) circle (2pt);
        \draw (4.969,-.171) circle (2pt);
        \filldraw [gray!50] (4.031,-.171) circle (2pt);
        \draw (4.031,-.171) circle (2pt);
        \end{tikzpicture}\\
        & -\frac{1}{4} ~
        \begin{tikzpicture}[baseline=-.1cm]
        \draw (6,0) circle (.5cm);
        \draw (6.5,0) to[out=180,in=90] (6,-.5);
        \filldraw [gray!50] (6,.5) circle (2pt);
        \draw(6,.5) circle (2pt);
        \filldraw [gray!50] (6.5,0) circle (2pt);
        \draw (6.5,0) circle (2pt);
        \draw (5.5,0) to[out=0,in=90] (6,-.5);
        \filldraw [gray!50] (5.5,0) circle (2pt);
        \draw (5.5,0) circle (2pt);
        \filldraw [gray!50] (6,-.5) circle (2pt);
        \draw(6,-.5) circle (2pt);
        \end{tikzpicture}
        ~ + \hspace{1pt} 2 \, ~
       \begin{tikzpicture}[baseline=-.1cm]
       \draw (0,0) circle (.5cm);
       \draw (-.5,0) to[out=0,in=180] (.5,0);
       \filldraw [gray!50] (0,.5) circle (2pt);
       \draw(0,.5) circle (2pt);
       \filldraw [gray!50] (.5,-0) circle (2pt);
       \draw (.5,0) circle (2pt);
       \draw (-.5,0) to[out=0,in=90] (0,-.5);
       \filldraw [gray!50] (-.5,0) circle (2pt);
       \draw (-.5,0) circle (2pt);
       \filldraw [gray!50] (0,-.5) circle (2pt);
       \draw(0,-.5) circle (2pt);
       \end{tikzpicture}
       \,-\frac{3}{16} ~
      \begin{tikzpicture}[baseline=-.1cm]
      \draw (1.5,0) circle (.5cm);
      \draw (1,0) to[out=-45,in=135] (1.5,-.5);
      \filldraw [gray!50] (1.5,.5) circle (2pt);
      \draw(1.5,.5) circle (2pt);
      \filldraw [gray!50] (2,0) circle (2pt);
      \draw (2,0) circle (2pt);
      \draw (1,0) to[out=0,in=90] (1.5,-.5);
      \filldraw [gray!50] (1,0) circle (2pt);
      \draw (1,0) circle (2pt);
      \filldraw [gray!50] (1.5,-.5) circle (2pt);
      \draw(1.5,-.5) circle (2pt);
      \end{tikzpicture}
      ~ + \hspace{1pt} \frac{1}{4} \hspace{2pt} ~
     \begin{tikzpicture}[baseline=-.1cm]
     \draw (3,0) circle (.5cm);
     \draw (2.5,0) to[out=30,in=150] (3.5,0);
     \filldraw [gray!50] (3,.5) circle (2pt);
     \draw(3,.5) circle (2pt);
     \draw (2.5,0) to[out=-30,in=210] (3.5,0);
     \filldraw [gray!50] (3.5,0) circle (2pt);
     \draw (3.5,0) circle (2pt);
     \filldraw [gray!50] (2.5,0) circle (2pt);
     \draw (2.5,0) circle (2pt);
     \filldraw [gray!50] (3,-.5) circle (2pt);
     \draw(3,-.5) circle (2pt);
     \end{tikzpicture} ~,\\
%
%     \beta_Z  = &
     \gamma_{ij}\phi_i\phi_j = &
     + \frac{1}{4} ~
     \begin{tikzpicture}[baseline=-.1cm]
     \draw (0,0) circle (.5cm);
     \draw (-.5,0) to [out=0,in=180] (.5,0);
     \filldraw [gray!50] (-.5,0) circle (2pt);
     \draw (-.5,0) circle (2pt);
     \filldraw [gray!50] (.5,0) circle (2pt);
     \draw(.5,0) circle (2pt);
     \end{tikzpicture}
     \, \, - \frac{1}{16} ~ \,
     \begin{tikzpicture}[baseline=-.1cm]
     \draw (0,0) circle (.5cm);
     \draw (0,.5) to [out=-90,in=30] (-.469,-.171);
     \draw (0,.5) to [out=-90,in=150] (.469,-.171);
     \filldraw [gray!50] (0,.5) circle (2pt);
     \draw (0,.5) circle (2pt);
     \filldraw [gray!50] (.469,-.171) circle (2pt);
     \draw (.469,-.171) circle (2pt);
     \filldraw [gray!50] (-.469,-.171) circle (2pt);
     \draw (-.469,-.171) circle (2pt);
     \end{tikzpicture} ~,\\
     A = &
      - \frac{1}{2} ~
      \begin{tikzpicture}[baseline=-.1cm]
        \draw (0,0) circle (.5cm);
        \draw (-0.5,0) to[out=30,in=150] (.5,0);
        \draw (-0.5,0) to[out=-30,in=210] (.5,0);
        \filldraw [green!50] (.5,0) circle (2pt);
        \draw (.5,0) circle (2pt);
        \filldraw [green!50] (-.5,0) circle (2pt);
        \draw (-.5,0) circle (2pt);
      \end{tikzpicture}
      ~ + ~
      \begin{tikzpicture}[baseline=-.1cm]
      \draw (1.5,0) circle (.5cm);
      \draw (1.5,.5) to [out=-90,in=30] (1.031,-.171);
      \draw (1.5,.5) to [out=-90,in=150] (1.969,-.171);
      \filldraw [green!50] (1.5,.5) circle (2pt);
      \draw (1.5,.5) circle (2pt);
      \draw (1.031,-.171) to[out=-20,in=200] (1.969,-.171);
      \filldraw [green!50] (1.969,-.171) circle (2pt);
      \draw (1.969,-.171) circle (2pt);
      \filldraw [green!50] (1.031,-.171) circle (2pt);
      \draw (1.031,-.171) circle (2pt);
      \end{tikzpicture}
      ~ - \frac{3}{2} ~
      \begin{tikzpicture}[baseline=-.1cm]
      \draw (3,0) circle (.5cm);
      \draw (3,.5) to[out=-90,in=180] (3.5,0);
      \draw (2.5,0) to[out=0,in=180] (3.5,0);
      \draw (3,.5) to[out=-90,in=90] (3,-.5);
      \filldraw [green!50] (3,.5) circle (2pt);
      \draw(3,.5) circle (2pt);
      \filldraw [green!50] (3.5,0) circle (2pt);
      \draw (3.5,0) circle (2pt);
      \draw (2.5,0) to[out=0,in=90] (3,-.5);
      \filldraw [green!50] (2.5,0) circle (2pt);
      \draw (2.5,0) circle (2pt);
      \filldraw [green!50] (3,-.5) circle (2pt);
      \draw(3,-.5) circle (2pt);
      \end{tikzpicture}
      ~ + \frac{1}{12} ~
      \begin{tikzpicture}[baseline=-.1cm]
      \draw (4.5,0) circle (.5cm);
      \draw (4,0) to[out=-45,in=135] (4.5,-.5);
      \draw (4.5,.5) to (5,0);
      \draw (4.5,.5) to[out=-90,in=180] (5,0);
      \filldraw [green!50] (4.5,.5) circle (2pt);
      \draw(4.5,.5) circle (2pt);
      \filldraw [green!50] (5,0) circle (2pt);
      \draw (5,0) circle (2pt);
      \draw (4,0) to[out=0,in=90] (4.5,-.5);
      \filldraw [green!50] (4,0) circle (2pt);
      \draw (4,0) circle (2pt);
      \filldraw [green!50] (4.5,-.5) circle (2pt);
      \draw(4.5,-.5) circle (2pt);
      \end{tikzpicture}~,
     \end{split}
\end{equation}

By plugging the effective potential \eqref{eq:pot4} into Eq.~\eqref{BBetasd=4} and switching to dimensionless variables, one can read off the corresponding $\beta$-functions $\beta_{m^2}$, $\beta_{\lambda_1}$ and $\beta_{\lambda_2}$, the renormalization group flow of the $A$ function and that of the anomalous dimension matrix $\gamma_{ij}$. The procedure is straightforward but requires some computational power.
They read\footnote{Hereafter by $\zeta_n$ we intend the Riemann zeta function.}
\begin{align}\label{eq:betasystem-dc4}
    \beta_{m^2} & = -2 m^2 +\frac{1}{3} m^2 [(N+2)\lambda_1 + 3 \lambda_2] - \frac{5}{18} m^2 \left[(N+2)\lambda_1^2 + 6 \lambda_1\lambda_2 + 3 \lambda_2^2 \right]\nonumber\\
    & + \frac{1}{72} m^2 \left[ 2 \left(5 N^2+47 N+74\right) \lambda_1^3 + 18 (5N+37) \lambda_1^2 \lambda_2 + 3 (N + 251) \lambda_1 \lambda_2^2 +252 \lambda_2^3 \right] \,,\\
    \beta_{\lambda_1} & = -\lambda_1 \epsilon + \frac{1}{3} \lambda_1 \left[(N+8) \lambda_1 + 6  \lambda_2 \right] - \frac{1}{3} \left[3 (N+14) \lambda_1^3 + 22 \lambda_1^2 \lambda_2 + 5 \lambda_1 \lambda_2^2\right] +7\lambda_1 \lambda_2^3 \nonumber\\
    & +\frac{1}{24} (N+192 \zeta_3+642) \lambda_1^2 \lambda_2^2  + \frac{1}{36} (79 N+768 \zeta_3 + 1318) \lambda_1^3 \lambda_2   \nonumber\\
    & +\frac{1}{216} \lambda_1^4 \left[33 N^2 + 922 N + (480 N +2112) \zeta_3+2960\right] \,,\\
    %%%%%
    \beta_{\lambda_2} & = -\lambda_2 \epsilon +4 \lambda_1 \lambda_2+3 \lambda_2^2 - \frac{1}{9} \left[(5 N + 82) \lambda_1^2 \lambda_2 + 138
    \lambda_1 \lambda_2^2 + 51 \lambda_2^3\right] +\frac{1}{2} \lambda_1 \lambda_2^3 (96 \zeta_3 + 131) \nonumber\\
    & +\frac{1}{24} \lambda_1^2 \lambda_2^2 (17
    N+1536 \zeta_3+1950) \!-\! \frac{1}{108} \lambda_1^3 \lambda_2 \left[13 N^2 -\! 368 N + 192 (N-14) \zeta_3 - 3284\right] \nonumber\\
    & + \frac{1}{8} \lambda_2^4 (96 \zeta_3+145) \,, \\
    \eta & = \frac{1}{18}(N \!+\! 2)\lambda_1^2  +\! \frac{1}{3} \lambda_1 \lambda_2 +\!  \frac{1}{6}\lambda_2^2 -\! \frac{1}{24} (N \!+\! 8) \lambda_1^2 \lambda_2
    -\frac{3}{8} \lambda_1 \lambda_2^2 - \frac{1}{8} \lambda_2^3 -\! \frac{1}{216} (N\!+\!2) (N\!+\!8) \lambda_1^3 \,,\\
    A & = -\frac{1}{6} N \left[(N+2) \,\lambda_1^2 + 6 \,\lambda_1 \lambda_2 + 3 \,\lambda_2^2\right] \,\epsilon
    + \frac{1}{27} N \left[ 9 (N+8) \,\lambda_1^2 \lambda_2 + 81\, \lambda_1 \lambda_2^2 + 27 \,\lambda_2^3 \right.\nonumber\\
    &\left. +\left(N^2+10 N+16\right)\lambda_1^3 \right]
    - \frac{1}{36} N \left[\left(3 N^2+20 N+28\right) \lambda_1^4 +12 (3 N+14) \lambda_1^3 \lambda_2 +204 \lambda_1 \lambda_2^3\right. \nonumber\\
    & \left. + 2 (5 N+148) \lambda_1^2 \lambda_2^2+51 \lambda_2^4\right] \,.
\end{align}
The gamma functions associated to the quadratic deformations are given by
\begin{align}
  \gamma_S
  & = -\frac{1}{216} \lambda_1^3 (N\!+\!2) (31 N\!+\!230)-\frac{1}{24} \lambda_1^2 \lambda_2 (31 N\!+\!230) + \frac{1}{3} \lambda_1^2 (N\!+\!2)-\frac{1}{24} \lambda_1 \lambda_2^2(N\!+\!260) \notag \\
  & + 2 \lambda_1 \lambda_2 - \frac{1}{3} \lambda_1 (N+2) - \frac{29\lambda_2^3}{8} + \lambda_2^2 - \lambda_2 \\
  \gamma_X
  & = \frac{1}{108} \lambda_1^3 \left(2 N^2\!-\!47 N\!-\!230\right)-\frac{5}{24} \lambda_1^2 \lambda_2 (N+46)+\frac{1}{9} \lambda_1^2 (N+6)-\frac{65 \lambda_1 \lambda_2^2}{6}+2
   \lambda_1 \lambda_2\notag\\
  & -\frac{2 \lambda_1}{3}  -\frac{29 \lambda_2^3}{8}+\lambda_2^2-\lambda_2 \,,\\
  \gamma_Y
  & = \frac{1}{108} \lambda_1^3 \left(2 N^2\!-\!47 N\!-\!230\right)+\frac{1}{6} \lambda_1^2 \lambda_2 (N-21)+\frac{1}{9} \lambda_1^2 (N+6)-\frac{7 \lambda_1
   \lambda_2^2}{12}+\frac{2 \lambda_1 \lambda_2}{3}-\frac{2 \lambda_1}{3}\,.
\end{align}

In the following, for each of the three nontrivial solutions to \eqref{eq:betasystem-dc4}, namely decoupled Ising, $O(N)$ and $H_N$, we report the relative NNLO fixed point coordinates, the anomalous dimension $\eta$, the critical exponents $\theta_R$ relative to the quadratic operators $R=S,X,Y$ and the fixed point value of the $A$-function (the latter, to our best knowledge, has been given at NLO in Ref.~\cite{Osborn:2017ucf}; here we provide the NNLO contribution).
We recall that the relation between the gamma function $\gamma_{\mathcal{R}}$, the critical exponent $\theta_{\mathcal{R}}$ and the (CFT) scaling dimension $\Delta_{\mathcal{R}}$ are given by the relations $\theta_R = d - \Delta_R$ and $\gamma_R$ as $\gamma_R = \theta_R - (d + 2 - \eta )/2$.
Finally, the correlation length critical exponent $\nu = \theta_S^{-1}$ and the crossover exponents are given by $\Phi_{\text{axis}} = \theta_X/\theta_S$ and $\Phi_{\text{diag}} = \theta_Y/\theta_S$.

\paragraph{Ising.}
The fixed point coordinates realtive to the Ising fixed point are given by
\begin{equation}
\begin{split}
  \lambda_1^\star & = 0 \,, \\
  \lambda_2^\star & = \frac{1}{3} \epsilon +\frac{17}{81} \epsilon ^2 + \frac{(709-2592 \zeta_3)}{17496}  \epsilon ^3 \,.
\end{split}
\end{equation}
The corresponding critical exponents and $A$ function read
\begin{equation}
\begin{split}
  \eta & =  \frac{\epsilon ^2}{54} + \frac{109 \epsilon ^3}{5832}\,, \\
  \theta_S & = 2 -\frac{1}{3}\epsilon - \frac{19}{162} \epsilon ^2 + \frac{(2592 \zeta_3 -937)}{17496}  \epsilon ^3 \,, \\
  \theta_X & = 2 -\frac{1}{3}\epsilon - \frac{19}{162} \epsilon ^2 + \frac{(2592 \zeta_3 -937)}{17496}  \epsilon ^3 \,, \\
  \theta_Y & = 2 -\frac{1}{54} \epsilon ^2 -\frac{109}{5832}  \epsilon ^3\,, \\
  A^\star & = - \frac{1}{54} \epsilon ^3 - \frac{17}{972} \epsilon ^4 \,.
\end{split}
\end{equation}

\paragraph{\texorpdfstring{\bm{$O(N)$}}{O(N)} model.}

The fixed point coordinates relative to the $O(N)$ point are given by
\begin{equation}\label{eq:fpONd4}
\begin{split}
  \lambda_1^\star & = \frac{3}{N+8} \epsilon + \frac{9 (3 N + 14)}{(N + 8)^3} \epsilon^2 \\
  & - \frac{99 N^3 + 30 N^2 (48 \zeta_3 - 11) + 96 N (186 \zeta_3-55) +192 (264 \zeta_3-71)}{8 (N+8)^5} \epsilon ^3  \,, \\
  \lambda_2^\star & = 0 \,.
\end{split}
\end{equation}
The corresponding critical exponents and $A$ function read
\begin{equation}
\begin{split}
  \eta & = \frac{(N+2)}{2 (N+8)^2}  \epsilon ^2 -\frac{(N+2) \left(N^2-56 N-272\right) }{8(N+8)^4} \epsilon ^3\,, \\
  \theta_S & = 2 - \frac{(N+2)}{N+8} \epsilon -\frac{(N+2) (13 N+44)}{2(N+8)^3} \epsilon^2 \\
  & + \frac{(N+2)\left[3 N^3 + N^2 (480 \zeta_3-452)+16 N (372 \zeta_3-167) + 64 (264 \zeta_3-83)\right]}{8 (N+8)^5}  \epsilon ^3\,, \\
  \theta_X & = 2 - \frac{2}{N+8}  \epsilon + \frac{\left(N^2-18 N-88\right)}{2(N+8)^3}  \epsilon ^2 \\
  & + \frac{\left[5 N^4+134 N^3+8 N^2 (120 \zeta_3-7)+32 N (372 \zeta_3-131)+128 (264 \zeta_3-83)\right]}{8 (N+8)^5} \epsilon ^3 \,, \\
  \theta_Y & = 2 - \frac{2}{N+8}  \epsilon + \frac{\left(N^2-18 N-88\right)}{2(N+8)^3}  \epsilon ^2 \\
  & + \frac{\left[5 N^4+134 N^3+8 N^2 (120 \zeta_3-7)+32 N (372 \zeta_3-131)+128 (264 \zeta_3-83)\right]}{8 (N+8)^5} \epsilon ^3 \,, \\
  A^\star & = -\frac{N (N+2)}{2 (N+8)^2} \epsilon ^3 - \frac{9 N \left(3 N^2+20 N+28\right)}{4 (N+8)^4}  \epsilon ^4 \,.
\end{split}
\end{equation}

\paragraph{\texorpdfstring{\bm{$H_N$}}{H_N} model.}

The fixed point with $H_N$ symmetry is the most interesting for the discussion of this paper.
The fixed point coordinates are given by
\begin{equation}\label{eq:fpHNd4}
\begin{split}
  \lambda_1^\star & = \frac{1}{N} \epsilon - \frac{\left(19 N^2 - 125 N + 106\right)}{27 N^3} \epsilon^2 + \frac{1}{5832 N^5} \left[N^4 (5184 \zeta_3 \! - \! 1955) \! - \! N^3 (15552 \zeta_3 + 41971)\right. \\
  & \left. - 6 N^2 (3024\zeta_3 - 38329) + 448 N (81 \zeta_3 - 805) + 179776\right] \epsilon ^3 \,, \\
  \lambda_2^\star & =  +\frac{(N-4)}{3 N} \epsilon  + \frac{\left(17 N^3+93 N^2-534 N+424\right)}{81N^3} \epsilon ^2 + \frac{1}{17496 N^5} \left[N^5 (709-2592 \zeta_3) \right. \\
  & \left. + N^4 (11713-20736 \zeta_3)+2 N^3 (25920 \zeta_3+90281)+N^2 (93312 \zeta_3-989656)\right. \\
  & \left. - 64 N (2268 \zeta_3-23441)-719104\right] \epsilon ^3  \,.
\end{split}
\end{equation}
and the critical exponents become
\begin{equation}
\begin{split}
  \eta & =  \frac{(N-1) (N+2)}{54 N^2} \epsilon ^2 + \frac{(N-1) \left(109 N^3-222 N^2+1728 N-1696\right)}{5832 N^4} \epsilon ^3 \\
  \theta_S & = 2  -\frac{2 (N-1)}{3 N}\epsilon + \frac{(N-1)\left(17 N^2-326 N+424\right)}{162 N^3} \epsilon ^2 - \frac{(N-1)}{17496 N^5} \left[N^4 (2592 \zeta_3-937) \right. \\
  & \left. - 2 N^3 (11664 \zeta_3+14179)-24 N^2 (864 \zeta_3-11039) + 32 N (2268 \zeta_3-17929)+359552\right]\epsilon ^3 \\
  \theta_X & = 2 - \frac{(N-2)}{3 N} \epsilon - \frac{\left(19 N^3+131 N^2-538 N+424\right)}{162 N^3} \epsilon ^2 + \frac{1}{17496 N^5}\left[N^5 (2592 \zeta_3-937)\right. \\
  & +3 N^4 (3456 \zeta_3-191)-2 N^3 (10368 \zeta_3 + 53629)- 88 N^2 (648 \zeta_3 - 5779) \\
  & \left. + 96 N (756 \zeta_3-7849)+359552\right] \epsilon ^3   \\
  \theta_Y & = 2 -\frac{2}{3 N} \epsilon + \frac{\left(-3 N^3+127 N^2-530 N+424\right)}{162 N^3}  \epsilon ^2 + \frac{1}{17496 N^5} \left[-327 N^5 \right. \\
  & + N^4 (2041-10368 \zeta_3) + N^3 (31104 \zeta_3 + 99170) + 216 N^2 (168\zeta_3 - 2291) \\
  & \left. + N (746720-72576 \zeta_3)-359552\right]\epsilon ^3 \,, \\
  A^\star & = -\frac{\left(N^2 + N-2\right) \epsilon^3}{54 N} - \frac{(N-1)^2 \left( 17 N^2 - 4N + 212 \right)\epsilon ^4}{972 N^3}\,.
\end{split}
\end{equation}

There are two important things to notice. The first one is that the limit $N\to 0$ of \eqref{eq:fpHNd4} is divergent, because of inverse powers of $N$.
The reason why this happens is well-known and it has to do with the fact that for $N=0$ the beta functions $\beta_{\lambda_1}$ and $\beta_{\lambda_1}$
have a degenerate solution, so it is necessary to consider a subleading equation when studying the $\epsilon$-expansion.
This, in turn, makes it such that the expansion is not analytic in $\epsilon$, but rather in $\epsilon^{\nicefrac{1}{2}}$, and the solution is known as
Khmelnitskii fixed point \cite{khmel1975}. Obviously, the unpleasant effect is that the expansion converges much less for finite values of $\epsilon$, but
it is still very important to see the fixed point because it has the physical meaning of describing an Ising model in a lattice that has random impurities
or that is randomly diluted, according to some simple distribution \cite{Grinstein:1976}.

\begin{figure}[h]
  {
  \center
  \includegraphics[width=.7\textwidth]{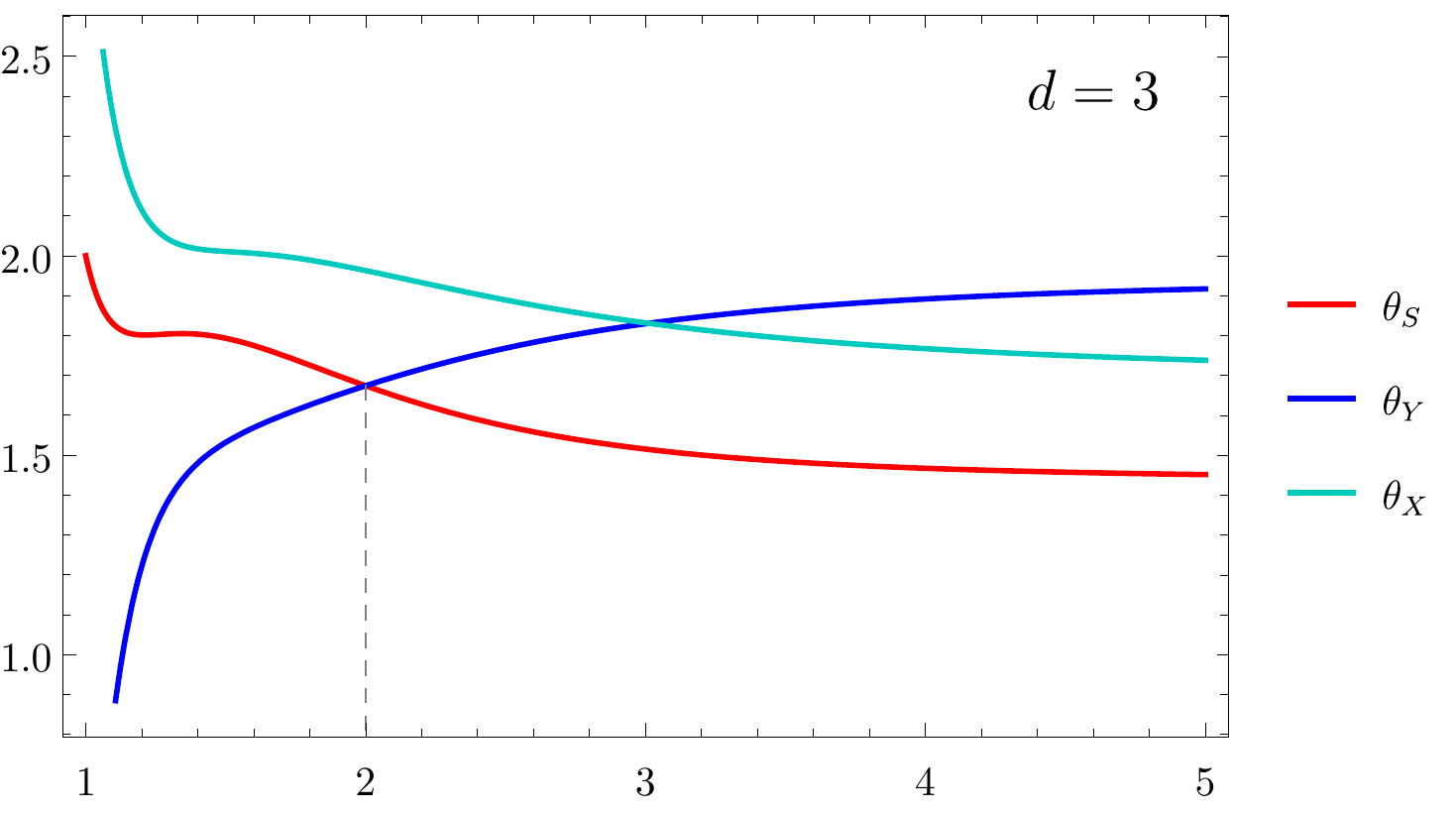}
  \caption{Plot of the critical exponents $\theta_R$ for $R=S,X,Y$ as a function of $N$, evaluated at the hypercubic fixed point
and extrapolated to $d=3$ ($\epsilon=1$). We see that at $N=2$, the singlet coincide with the diagonal quadratic deformation. This is expected since at $N=2$ the cubic behavior reduces to two decoupled Ising models and the diagonal deformation simply amounts at a shift in the critical temperature (this holds true at all orders in perturbation theory). At $N\approx 3$ the axial and diagonal quadratic deformations almost coincide. This is expected too since at $N=N_c$ the cubic and isotropic fixed points coincide and the corresponding two crossovers are equivalent, and $N_c$ is known to be close to $3$.
Notice, however, that this extrapolation gives $N_c \gtrsim 3$, while it is believed that $N_c\simeq 2.9 \lesssim 3$, and the estimate is fixed by a careful resummation. The analytical expression of the crossover exponents $\Phi_{\text{axis}}$ and $\Phi_{\text{diag}}$ can be checked at order $O(\epsilon^2)$ against Ref.~\cite{AHARONY1976}. \label{fig:exponents}}
  }
\end{figure}

The second important point can be made in relation to the stability and $N_c$ at $d=3$. Specifically, as seen from Fig.~\ref{fig:exponents}
we notice that the critical exponents of the $X$ and $Y$ quadratic operators of the irreps of $H_N$ almost coincide for $N=3$, implying that
axial and quadratic deformations scale very similarly in that configuration.
This has to be compared with the analysis of the irreps of the $O(N)$ model, which contain the same singlet $S$ as the hypercubic model, but
combine the $X$ and $Y$ operators in a single traceless tensor with two indices. This implies that if $\gamma_X=\gamma_Y$
then the hypercubic model must coincide with the $O(N)$ model, so, from our extrapolation to $\epsilon=1$
we deduce that for $N=d=3$ the two models must be very close in some sense.
In fact, one can check explicitly that the perturbative solution of $\gamma_X=\gamma_Y$, that gives $N_c=N_c(\epsilon)$ in $d=4-\epsilon$,
has the same solution as the condition for which the two fixed points \eqref{eq:fpONd4} and \eqref{eq:fpHNd4} coincide. It is actually
even simpler to determine $N_c(\epsilon)$ in this way.

%Finally, the last quantity we compute is $C_T$ which is rooted in the CFT formalism and is obtained when expanding the correlator involving the trace of the stress-energy tensor $T$ and two copies of the field $\phi_i$. {\color{blue} Add here the formula to compute it from $V$ from Ref.~\cite{Dey:2016}}
%
%\begin{align}
%  C_T  & = N -\frac{5}{108} N \left[\lambda_1^2 (N+2)+6 \lambda_1 \lambda_2+3\lambda_2^2\right]\nonumber\\
% & - \frac{7}{972} N \left[\left(N^2+10N+16\right) \lambda_1^3 + 9 (N+8) \lambda_1^2 \lambda_2 + 81 \lambda_1 \lambda_2^2 + 27 \lambda_2^3\right] \,.
%\end{align}

\section{The model with quintic interaction in \texorpdfstring{\bm{$d = \frac{10}{3}-\epsilon$}}{10/3-eps}}
\label{sec:10/3-eps}

In this Section we report the system of $\beta$-functions relevant to the study of the hypercubic theory in $d_c=\frac{10}{3}$, which was
discussed in \cite{Zinati:2020xcn}. We do so for completeness. The RG flow can be obtained at LO by promoting the single-field case studied in Ref.~\cite{Codello:2017epp}
to an arbitrary number of flavors. The resulting beta functionals are given by
\begin{equation}\label{eq:rg10over3}
\begin{split}
        \beta_V  = &
        + \, \frac{3}{4} ~ \, \,
        \begin{tikzpicture}[baseline=-.1cm]
        \draw (0,0) circle (.5cm);
        \draw (-.5,0) to [out=30,in=150] (.5,0);
        \draw (-.5,0) to[out=-30,in=-150] (.5,0);
        \filldraw [gray!50] (0,.5) circle (2pt);
        \draw (0,.5) circle (2pt);
        \filldraw [gray!50] (.5,0) circle (2pt);
        \draw (.5,0) circle (2pt);
        \filldraw [gray!50] (-.5,0) circle (2pt);
        \draw (-.5,0) circle (2pt);
        \end{tikzpicture}
        ~ - \frac{27}{8} ~
        \begin{tikzpicture}[baseline=-.1cm]
        \draw (1.5,0) circle (.5cm);
        \draw (1.5,.5) to[out=-90,in=30] (1.067,-.25);
        \draw (1.5,.5) to[out=-90,in=150] (1.933,-.25);
        \filldraw [gray!50] (1.5,.5) circle (2pt);
        \draw(1.5,.5) circle (2pt);
        \filldraw [gray!50] (1.933,-.25) circle (2pt);
        \draw (1.933,-.25) circle (2pt);
        \filldraw [gray!50] (1.067,-.25) circle (2pt);
        \draw (1.067,-.25) circle (2pt);
        \end{tikzpicture}\\
%
%     \beta_Z  = &
       \gamma_{ij}\phi_i\phi_j = &
        + \frac{3}{80} ~
        \begin{tikzpicture}[baseline=-.1cm]
        \draw (0,0) circle (.5cm);
        \draw (-.5,0) to [out=30,in=150] (.5,0);
        \draw (-.5,0) to[out=-30,in=-150] (.5,0);
        \filldraw [gray!50] (.5,0) circle (2pt);
        \draw (.5,0) circle (2pt);
        \filldraw [gray!50] (-.5,0) circle (2pt);
        \draw (-.5,0) circle (2pt);
        \end{tikzpicture} \,.
     \end{split}
\end{equation}

The inclusion of a singlet $\psi$ in the multiplet of fields $\phi_i \to (\phi_i,\psi)$ allows for the extension of a theory with $H_N$ symmetry
to an odd interaction (an odd interaction would otherwise break $H_N$). The first possible interaction is the cubic in $d=6-\epsilon$,
but the singlet $\psi$ could only couple to the singlet $(\phi_i)^2$, which is also $O(N)$ invariant, so it would reduce to the $O(N)$ theory studied in Ref.~\cite{Giombi:2019}.
The first genuinely hypercubic model comes from a quintic interaction, which can be expressed in terms of the following effective potential
\begin{align}
  V(\phi_i, \psi) & = \kappa_1 \psi^5 + \kappa_2 \psi^3 \sum_i \phi_i^2 + \kappa_3 \psi \Bigl(\sum_i \phi_i^2\Bigr)^2 + \kappa_4 \psi \sum_i \phi_i^4 \nonumber \\
  & + \omega_1 \psi^3 + \omega_2 \psi \sum_i \phi_i^2 + m_1 \psi^2 + m_2 \sum_i \phi_i^2 \,,
\end{align}
and therefore follows the RG of \eqref{eq:rg10over3}.

The corresponding system of $\beta$-functions of the quadratic singlest are
\begin{align}
  \beta_{m^2_1} & = -2 m^2_1 +\frac{1}{1125} N^2 (m^2_1+10 m^2_2) \kappa_3^2  + \frac{1}{1125} N \left(27 \kappa_2^2+2 \kappa_3^2+6 \kappa_3 \kappa_4+3
   \kappa_4^2\right)  m^2_1 \nonumber\\
   & + \frac{1}{450} N \left(9 \kappa_2^2+8 \kappa_3^2+24 \kappa_3
    \kappa_4+12 \kappa_4^2\right) m^2_2  +\frac{11}{15} m^2_1 \kappa_1^2\,,\\
  \beta_{m^2_2} & = -2 m^2_2 +\frac{2}{1125} N (5 m^2_1+17 m^2_2)\kappa_3^2 + \frac{1}{450} \left(9 \kappa_2^2+8 \kappa_3^2+24 \kappa_3 \kappa_4+12\kappa_4^2\right) m^2_1 \nonumber\\
  & + \frac{1}{2250}\left(21 \kappa_2^2+ 136 \kappa_3^2 + 408 \kappa_3 \kappa_4 + 204 \kappa_4^2\right) m^2_2\,.
\end{align}
Cubic interactions flow as
\begin{align}
  \beta_{\omega_1} & = -\frac{1}{6}(8+3 \epsilon) \omega_1 +\frac{1}{2250} N^2
  \left(3 \omega_1 \kappa_3 + 20 \omega_2 \kappa_3 - 45 \omega_2 \kappa_2\right) \kappa_3
  -\frac{69}{10} \kappa_1^2 \omega_1
  \nonumber\\
  &
  + \frac{1}{450} N \left(135 \kappa_1 \kappa_2+45
   \kappa_2^2 + 18 \kappa_2\kappa_3 + 27 \kappa_2\kappa_4 - 8 \kappa_3^2 - 24
   \kappa_3 \kappa_4 - 12 \kappa_4^2 \right) \omega_2 \nonumber \\
  &
  - \frac{1}{750} N  \left(18 \kappa_2^2 - 2 \kappa_3^2 - 6 \kappa_3
   \kappa_4 - 3 \kappa_4^2\right)  \omega_1 \,,\\
  \beta_{\omega_2} & =
  -\frac{1}{6} (8 + 3 \epsilon) \omega_2
  \!-\!\frac{29}{2250} N^2 \kappa_3^2 \omega_2
  \!-\! \frac{1}{2250} N \left(63  \kappa_2^2+300 \kappa_2 \kappa_3\!+\!370 \kappa_3^2+174 \kappa_3 \kappa_4-3\kappa_4^2\right) \omega_2   \nonumber\\
  &
   + \frac{1}{750} \left(25 \kappa_1^2+100 \kappa_1 \kappa_2-53
   \kappa_2^2- 200 \kappa_2\kappa_3 - 300 \kappa_2\kappa_4 -208 \kappa_3^2 - 624
   \kappa_3 \kappa_4 - 342 \kappa_4^2 \right) \omega_2 \nonumber\\
   &
   -\frac{1}{150}  \left(135 \kappa_1
   \kappa_2+45 \kappa_2^2+ 18 \kappa_2 \kappa_3 + 27 \kappa_2 \kappa_4 -8
   \kappa_3^2 - 24 \kappa_3 \kappa_4 - 12 \kappa_4^2 \right)  \omega_1 \nonumber \\
   &
   +\frac{1}{150} N (4 \kappa_3-9 \kappa_2) \kappa_3 \omega_1\,.
\end{align}
Finally, the marginal interactions have beta functions
\begin{align}
  \beta_{\kappa_1} & = -\frac{3}{2} \epsilon \kappa_1 \!+\! \frac{1}{900} N^2
  \left(2 \kappa_1 \kappa_3-27 \kappa_2^2+8 \kappa_2
  \kappa_3\right) \kappa_3 \!-\! \frac{1}{900} N \left(621 \kappa_2^2 - 4
   \kappa_3^2 - 12 \kappa_3 \kappa_4 - 6 \kappa_4^2\right) \kappa_1 \nonumber\\
  &
  -\frac{1}{900} N \left(144
   \kappa_2^2 + 54 \kappa_2 \kappa_3 + 81 \kappa_2 \kappa_4 - 16 \kappa_3^2 + -48
   \kappa_3 \kappa_4 -24 \kappa_4^2 \right) \kappa_2  -\frac{229}{6}  \kappa_1^3\,,
 \end{align}
 \begin{align}
  \beta_{\kappa_2} & = -\frac{3}{2} \epsilon \kappa_2 -\frac{1}{2250} N^2 (87 \kappa_2+80
  \kappa_3)\kappa_3^2 - \frac{1}{2250} N \left(1350 \kappa_1 \kappa_2 \kappa_3 - 200 \kappa_1 \kappa_3^2 + 54
   \kappa_2^3 \right.\nonumber\\
   &\left.+ 1350 \kappa_2^2 \kappa_3 + 1246 \kappa_2 \kappa_3^2 + 522 \kappa_2
   \kappa_3 \kappa_4 - 9 \kappa_2 \kappa_4^2 + 1040 \kappa_3^3 + 720 \kappa_3^2
   \kappa_4 - 120 \kappa_3 \kappa_4^2\right) \nonumber\\
   &
   -\frac{1}{2250}\left(15525 \kappa_1^2 \kappa_2 + 10800 \kappa_1 \kappa_2^2 + 2700 \kappa_1
   \kappa_2 \kappa_3 + 4050 \kappa_1 \kappa_2 \kappa_4 - 400 \kappa_1
   \kappa_3^2 - 1200 \kappa_1 \kappa_3 \kappa_4 \right.\nonumber\\
   &  \vphantom{\frac{1}{2}}- 600 \kappa_1 \kappa_4^2 + 2139
   \kappa_2^3 + 2700 \kappa_2^2 \kappa_3 + 4050 \kappa_2^2 \kappa_4 + 2144 \kappa_2
   \kappa_3^2 + 6432 \kappa_2 \kappa_3 \kappa_4 + 3486 \kappa_2 \kappa_4^2  \nonumber\\
   &\vphantom{\frac{1}{2}}\left. + 1760
   \kappa_3^3 + 7920 \kappa_3^2 \kappa_4 + 8760 \kappa_3 \kappa_4^2 + 2880
   \kappa_4^3\right)\,,
 \end{align}
 \begin{align}
  \beta_{\kappa_3} & = -\frac{3}{2} \epsilon \kappa_3 -\frac{29}{2250} N^2 \kappa_3^3 -\frac{1}{4500} N \left(261 \kappa_2^2+720
  \kappa_2 \kappa_3+2164 \kappa_3^2+528 \kappa_3 \kappa_4-6
  \kappa_4^2\right) \kappa_3 \nonumber\\
  &
  + \frac{1}{4500}\left(150 \kappa_1^2 \kappa_3-2025 \kappa_1 \kappa_2^2+1200 \kappa_1 \kappa_2
   \kappa_3-1350 \kappa_2^3-3216 \kappa_2^2 \kappa_3-810 \kappa_2^2
   \kappa_4-7920 \kappa_2 \kappa_3^2 \right.\nonumber\\
   & \vphantom{\frac{1}{2}}\left. -5040 \kappa_2 \kappa_3 \kappa_4+360
   \kappa_2 \kappa_4^2-9616 \kappa_3^3-15408 \kappa_3^2 \kappa_4-4644
   \kappa_3 \kappa_4^2\right)\,,
 \end{align}
 \begin{align}
  \beta_{\kappa_4} & = -\frac{3}{2} \epsilon \kappa_4 +\frac{1}{2250} N^2 \kappa_3^2 \kappa_4
  +\frac{1}{4500} N
  \left(9 \kappa_2^2+240 \kappa_2 \kappa_3-1004 \kappa_3^2+12 \kappa_3
  \kappa_4+6 \kappa_4^2\right) \kappa_4 \nonumber\\
  & +\frac{1}{750}\!\left(25 \kappa_1^2\!+\!200 \kappa_1 \kappa_2\!-\!446
   \kappa_2^2\!-\!2080 \kappa_2 \kappa_3\!-\!1500 \kappa_2 \kappa_4\!-\!3096
   \kappa_3^2\!-\!5148 \kappa_3 \kappa_4\!-\!1974 \kappa_4^2\right)  \kappa_4
\end{align}
In the limit $N\to 0$ they reduce to the system of $\beta$-functions given in Ref.~\cite{Zinati:2020xcn}.

\bibliographystyle{jhep.bst}
\bibliography{mchc}

\end{document}